\pdfoutput=1
\documentclass[aps,pra,twocolumn,showpacs,superscriptaddress,10pt,letterpaper,floatfix]{revtex4-1}

\usepackage[utf8]{inputenc} 
\usepackage[T1]{fontenc}	
\usepackage{geometry}		
\usepackage{graphicx}
\usepackage[above,below]{placeins}	
\usepackage{times}
\usepackage{multirow}
\usepackage{hyperref}
\usepackage{amsmath}
\usepackage{cleveref}
\usepackage{enumitem}

\usepackage{hyperref}  
\hypersetup{colorlinks=true,urlcolor=blue,citecolor=blue,linkcolor=blue}   
\usepackage{enumitem}          
\setlist{nosep}                 

\begin{document}


  \title{The Physics GRE does not help applicants ``stand out''}

  \author{Nicholas T. Young}
  \affiliation {Department of Physics and Astronomy, Michigan State University, East Lansing, Michigan 48824}
  \affiliation {Department of Computational Mathematics, Science, and Engineering, Michigan State University, East Lansing, Michigan 48824}
    
  \author{Marcos D. Caballero}
  \affiliation {Department of Physics and Astronomy, Michigan State University, East Lansing, Michigan 48824}
  \affiliation {Department of Computational Mathematics, Science, and Engineering, Michigan State University, East Lansing, Michigan 48824}
  \affiliation {Center for Computing in Science Education \& Department of Physics, University of Oslo, N-0316 Oslo, Norway}
  \affiliation {CREATE for STEM Institute, Michigan State University, East Lansing, Michigan 48824}


  \begin{abstract}
   One argument for keeping the physics GRE is that it can help applicants who might otherwise be missed in the admissions process stand out. In this work, we evaluate whether this claim is supported by physics graduate school admissions decisions. We used admissions data from five PhD-granting physics departments over a 2-year period (N=2537) to see how the fraction of applicants admitted varied based on their physics GRE scores. We compared applicants with low GPAs to applicants with higher GPAs, applicants from large undergraduate universities to applicants from smaller undergraduate universities, and applicants from selective undergraduate institutions to applicants from less selective undergraduate institutions. We also performed a mediation and moderation analysis to provide statistical rigor and to better understand the previous relationships. We find that for applicants who might otherwise have been missed (e.g. have a low GPA or attended a small or less selective school) having a high physics GRE score did not seem to increase the applicant's chances of being admitted to the schools. However, having a low physics GRE score seemed to penalize otherwise competitive applicants. Thus, our work suggests that the physics GRE does not, in fact, help applicants who might otherwise be missed stand out.
  \end{abstract}

  \maketitle


\section{Introduction}
While applying to graduate programs requires many components, perhaps none is as scrutinized as the Graduate Records Exam (GRE), and in physics, the physics GRE. Indeed, research into graduate admissions in physics suggests that the physics GRE is one of the most important components of the applications for determining which applicants will be admitted, based on both student and faculty perspectives \cite{chari_understanding_2019, potvin_investigating_2017} and analysis of the admissions process \cite{,posselt_inside_2016, young_using_2020}. Despite its prominence in the admissions process, the physics GRE is known to be biased against women and people of color in physics \cite{miller_test_2014}, resulting in lower average scores compared to white and Asian males. At least one in three programs use a cutoff score \cite{potvin_investigating_2017}, with 700 being a common choice \cite{miller_typical_2019}, meaning applicants from groups already underrepresented in physics graduate programs can be further marginalized as they are less likely to achieve these scores. This is in addition to the observation that many physics students of color already see the GRE as a barrier to applying to graduate school \cite{cochran_identifying_2018,wilson_predicting_2020,owens_not_2020}.

Further, the physics GRE might not even be useful for determining which applicants will be successful in graduate school. For example, Miller et al. suggest that the physics GRE is not useful for predicting which applicants will earn their PhDs \cite{miller_typical_2019}. Additionally, Levesque et al. argue that using the common 50th percentile cutoff score for the physics GRE would have caused admissions committees to reject nearly 30\% of students who would later receive a national prize postdoctoral fellowship, which can be viewed as a proxy for research excellence \cite{levesque_physics_2015}. Yet despite evidence suggesting the physics GRE does not predict these typical ways of measuring ``success'' in graduate school and calls from the American Astronomical Society and the American Association of Physics Teachers to eliminate the physics GRE from admissions \cite{noauthor_statement_2019,noauthor_aas_2018}, most physics graduate programs still require applicants to submit their physics GRE scores. Currently, nearly 90\% of physics and astronomy graduate programs still accept the physics GRE, with over half requiring or recommending submitting a score \cite{noauthor_statement_2019}. Of those that do not accept physics GRE scores from applicants, all of the programs are solely astronomy graduate programs or joint physics and astronomy graduate programs. While it is uncertain where removing the physics GRE affects any measure of graduate school success (e.g. completion rate), initial work by Lopez suggests that removing the physics GRE does increase the diversity of applicants \cite{lopez_demographic_2019}.

Given these documented issues with the physics GRE, why do departments continue to use it? First, given that many programs are seeing larger number of applicants, the physics GRE provides a quick way to filter the applications down to a more reasonable number for faculty review. Unlike in undergraduate admissions, graduate admissions tend to be decentralized and done at the departmental level by a faculty committee. Hence, faculty are asked to review applications in addition to their regular teaching and research duties and thus, might not have the time to read the letters of recommendation and applicant essays for every applicant.

Second, some faculty view GRE scores as measures of innate intelligence \cite{posselt_inside_2016,scherr_fixed_2017} or ability to become a PhD-level scientist \cite{miller_test_2014}. After all, they and other faculty likely had high GRE scores in order to be admitted to graduate school, and may exhibit a survivorship bias, believing that a high GRE score is needed to succeed. Further, physics is seen as a "brilliance-required" field, where innate intelligence is required for success \cite{leslie_expectations_2015}.

A third argument, and the most interesting one in terms of the scope of this paper, is that standardized tests such as the physics GRE can help students stand out \cite{langin_wave_2019}. The ETS, the creator of the GRE and physics GRE, claims that subject GREs "can help you stand out from other applicants by emphasizing your knowledge and skill level in a specific area" \cite{noauthor_about_nodate}. For example, a student with an average grade point average (GPA) might be able to stand out from other applicants if they did exceptionally well on the physics GRE. 

In addition, applicants from smaller universities or universities that are not known to the admissions committee might benefit from performing well on a standardized measure. For example, the ETS claims that the GRE provide a ``common, objective measure to help programs compare students from different backgrounds'' \cite{noauthor_gre_2019} and physics admissions committees worry that removing the GRE would limit their ability to compare applicants from different backgrounds \cite{morrison_women_2020}. Anecdotally, some faculty claim that a good physics GRE score could aid students from small liberal arts colleges in the admissions process \cite{levesque_why_2017}.

We already know that GPAs are interpreted in context of the applicant's university. Posselt has shown that among more prestigious graduate programs, the applicant's GPA is viewed in the context of their undergraduate institution with high GPAs from prestigious institutions seen favorably, low GPAs from an unknown school as unfavorably, and high GPAs from unknown schools and middle GPAs from prestigious institutions in the middle \cite{posselt_trust_2018}. Therefore, a standardized test such as the physics GRE could provide an assumed equal comparison for an admissions committee and might allow the applicant from an unknown school to stand out or have a similar chance of admission as an applicant from a more well known school.

Finally, graduate admissions have been documented to be "risk-adverse," where admissions committees select applicants most likely to complete their program \cite{posselt_inside_2016,scherr_fixed_2017}. As applicants from smaller universities may be judged based on how previously enrolled students from their university did in the program \cite{posselt_trust_2018}, a risk adverse admissions committee might be less likely to admit applicants from small universities whose students have previously struggled in their program. However, perhaps a high standardized test score could overcome these perceptions and signal that the applicant might indeed be successful in the program.

Our goal then is to focus on the third argument. Does the physics GRE help applicants "stand out" in the admissions process? If that is the case, we would expect those disadvantaged in the admissions process, those who have low GPAs, attended a smaller institution, or identify as part of a group currently underrepresented in  physics, to be admitted at similar rates as their more advantaged peers with similar physics GRE scores. Specifically, we ask:

\begin{enumerate}
    \item How does an applicant's physics GRE score and undergraduate GPA affect their probability of admission?
    \item How are these probabilities of admission affected by an applicant's undergraduate institution, gender, and race?
\end{enumerate}

As Small points out in his critique of admissions and standardized test studies \cite{small_range_2017}, multiple variables rather than just a standardized test might best explain our results and therefore, a framework that allows for substitutions and trade-offs between variables is necessary. Therefore, we ask an additional research question:
\begin{enumerate}[resume]
    \item How might the above relationships be accounted for through mediating and moderating relationships?
\end{enumerate}

This paper is organized as follows: Sec. \ref{sec:Background} provides an overview of mediation and moderation analysis. We then describe our data, how we determined what constitutes ``standing out," and how we implemented mediation and moderation analysis in Sec. \ref{sec:methods}. In Sec. \ref{sec:Results}, we describe our findings and in Sec. \ref{sec:Discussion}, we use those findings to answer our research questions and explain our limitations and choices which may affect our results. Finally, we describe our future work in Sec. \ref{sec:FutureWork} and the implications of our work for graduate admissions in physics in Sec. \ref{sec:Conclusion}.

\section{Background\label{sec:Background}}
Before we can answer the third research question, it is important to describe what we mean by mediating and moderating relationships. 

In a mediating relationship, two variables are only related because they are also related to some common third variable. For example, a student who played video games the night before an exam might to do poorly because they stayed up playing video games too late and did not get enough sleep. Therefore, video games and doing poorly on the exam are only related due the common factor of lack of sleep. Lack of sleep is then a mediating variable.

In a moderating relationship, the strength of the relationship between two variables depends on some third variable. For example, the relationship between someone liking dogs and owning a dog likely depends on whether they are allergic to dogs. That is, we would expect someone who likes dogs but is allergic to dogs is less likely to own a dog than someone who likes dogs but is not allergic to dogs is. Being allergic to dogs is then a moderating variable.

\begin{figure}
 \includegraphics[width=1\linewidth]{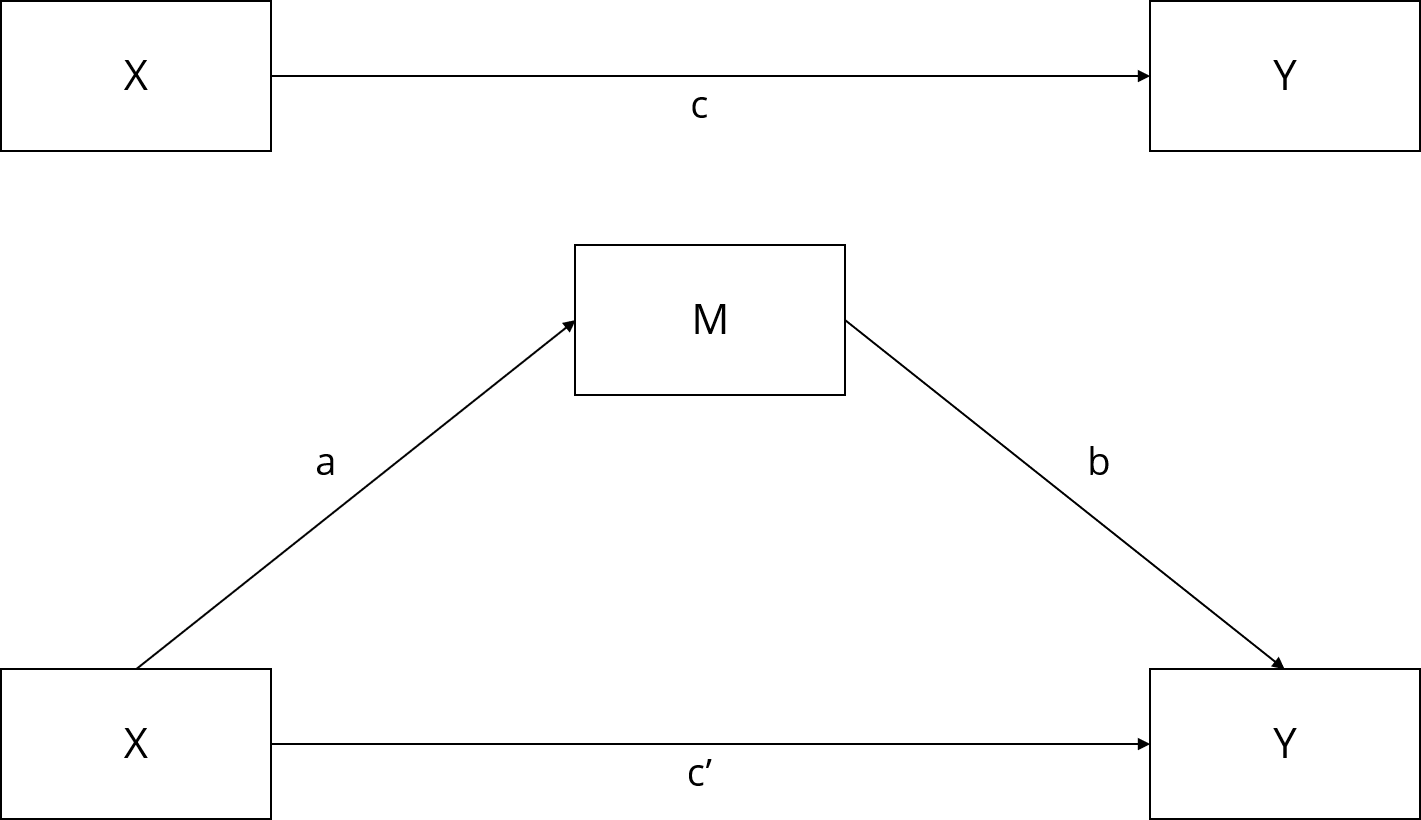}
 \caption{Visual representation of \cref{eqn1,eqn2,eqn3}. The top graphic shows \cref{eqn1} while the bottom graphic shows \cref{eqn2,eqn3}.\label{med_fig}}
\end{figure}

Mathematically, suppose that some input $X$ has an effect on output $Y$. We would say that some other input $M$ mediates the relationship between $X$ and $Y$ if $X$ only has an effect on $Y$ because $X$ has an effect on $M$ and $M$ has an effect on $Y$ \cite{baron_moderator-mediator_1986}. For a simple case, we can represent these relationships as

\begin{equation} \label{eqn1}
    Y=i_1 +cX
\end{equation}
\begin{equation} \label{eqn2}
    M=i_2 + aX
\end{equation}
\begin{equation} \label{eqn3}
    Y=i_3 + c'X+bM
\end{equation}

where $i$ represents the intercepts. These relationships are visually shown in Fig. \ref{med_fig}.

Using this representation, the direct effect of $X$ on $Y$ is represented by $c'$ and the indirect effect is represented by $ab$. The total effect is then $c'+ab$ which for a linear regression models, is equal to $c$. Equivalently, in the case the linear regression, the indirect effect is $c-c'$.

However, if $Y$ is binary, linear regression is not appropriate and logistic regression should be used instead. In this case, Rijnhart et al. recommend using $ab$ as the indirect effect as their simulation studies found the $ab$ estimate of the indirect effect exhibited less bias than the $c-c'$ estimate \cite{rijnhart_comparison_2019}.

To determine if the indirect effect is statistically significant, a common approach is to use a Sobel test. However, simulations suggest that the Sobel test is underpowered and that bootstrapping is a good alternative \cite{hayes_relative_2013}. Specifically, those simulations find that using the percentiles of a bootstrapped estimate of the indirect effect to estimate the confidence interval is a good compromise between avoiding type I errors while maintaining statistical power. From their approach (which has also been used in PER studies before, e.g. \cite{amos_mediating_2018}), if $ab$ is different than zero, then there is some degree of mediation.

More specifically, there are three cases.
\begin{enumerate}
    \item If $ab \neq 0$ and $c'=0$ then $M$ fully mediates the relationship between $X$ and $Y$.
    \item If $ab \neq 0$ and $c' \neq 0$, then $M$ partially mediates the relationship between $X$ and $Y$. In that case, we can estimate the amount of mediation as the fraction of the total effect attributed to the indirect effect, $\frac{ab}{ab+c'}$ \cite{ditlevsen_mediation_2005,freedman_confidence_2001}.
    \item If $ab=0$, then $M$ does not mediate the relationship between $X$ and $Y$.
\end{enumerate}

So far, we've assumed that the relationship between the mediator $M$ and the output $Y$ does not depend on any other variables. However, it is possible that the relationship between $M$ and $Y$ could also depend on $X$ or some other variable, meaning there is a conditional indirect effect (see Preacher et al. \cite{preacher_addressing_2007}). In the case that the relationship between $M$ and $Y$ depends on $X$, we would say that $X$ moderates the relationship between $M$ and $Y$. Practically, this means we must add an interaction term to \cref{eqn3}, which then becomes \cite{preacher_addressing_2007}

\begin{equation} \label{eqn4}
\begin{aligned}
    Y=i_3 + c'X+b_1M+b_2XM \\
     = i_3+c'X+(b_1+b_2X)M
\end{aligned}
\end{equation}.

The conditional indirect effect is then $a(b_1+b_2X)$. If $b_2=0$, we would say that there is no moderation and the indirect effect is the standard $ab$.

In the special case that $X$ is binary, \cref{eqn4} reduces to $Y=i_{x=0}+b_1M$ when $X=0$ and $Y=i_{X=1}+(b_1+b_2)M$ when $X=1$. Therefore, to test if there is moderation, we can simply regress $M$ on $Y$ given $X=0$ and again given $X=1$ and subtract the slopes to calculate $b_2$.

\section{Methods}\label{sec:methods}
\subsection{Data}
Data for this study comes from the physics departments at five selective, research-intensive, primarily white universities. Four of these universities are public and part of the Big Ten Academic Alliance while the remaining university is a private Midwestern university. During the 2017 and 2018 academic years, graduate admissions committees at these five universities recorded all physics applicants' undergraduate GPA, GRE scores, undergraduate institution, and demographic information such as gender, race, and domestic status. In addition, the universities recorded whether each applicant made the shortlist, was offered admission, and whether the applicant decided to enroll. Because our study includes all applicants rather than only admitted applicants, we are unlikely to suffer from the range restrictions noted in critiques of other admissions studies (e.g. \cite{small_range_2017,weissman_gre_2020}). However, we do a address a possible range restriction in the Limitations and Researcher Decisions section (sec. \ref{sec:limitations}).

Due to different requirements and admissions processes for international students and domestic students (e.g. international students need to submit a test of English proficiency), we only include domestic students in our study. We then remove any applicant for whom a physics GRE and GPA were not recorded, leaving us with 2537 applicants. Distributions and analysis of the physics GRE scores and GPAs appear in the appendix.

As the applicant's undergraduate university does not contain meaning in itself, we needed to categorize the institutions. We chose to categorize the institutions by their size and their selectivity. We then used the number of physics bachelor's degrees awarded per year as measure of the size of the university. We assume that universities with more graduates are more well known and hence, would likely be known to the admissions committees. In contrast, universities that produce fewer bachelor's degrees might not be known to the admissions committees and hence, might be unknown programs. It would then be these applicants from ``smaller'' programs who might need to ``stand out.'' We acknowledge some programs that produce a small number of physics bachelor's degrees each year might not be unknown to the admissions committees due to previous applicants from such schools or research collaborations or partnerships. However, there is no way in our data to know if this is the case.

To determine whether a university should be counted as a ``small university'', we used the undergraduate institution names to look up the number of typical physics bachelor's degrees from AIP's public degree data \cite{nicholson_roster_2018,nicholson_roster_2019}. As of this writing, degree data for the 2018 academic year was not available, so we used data from the 2016 and 2017 academic years to quantify the number of bachelor's degrees. Additionally, this would have been the most recent data available when admissions committees would have reviewed applications and many of the applicants would be represented in the data as bachelor degree recipients

To account for the institution's prestige, we used Barron's Selectivity Index \cite{national_center_for_education_statistics_nces-barrons_2017}. Barron's selectivity index is a measure based on the undergraduate acceptance rate of an institution as well as characteristics of its undergraduate incoming classes, such as mean SAT scores, high school GPAs, and class rank. We assume selectivity is a proxy for prestige as prestigious institutions tend to have low acceptance rates and high SAT scores and GPAs from incoming students. In contrast to the AIP data, Barron's selectivity index applies to the institution as a whole rather than only the physics department.

\subsection{Probability of admission procedure}

Determining whether an applicant is more or less likely to be admitted first requires computing admissions probabilities. To do so, we grouped applicants based on their GPAs and physics GRE scores. Prior work has found that the physics GRE score and undergraduate GPA are two of the most important aspects of the applications \cite{posselt_inside_2016,potvin_investigating_2017,young_using_2020}. Our previous work specifically found that the physics GRE score and undergraduate GPA were able to predict with 75\% accuracy whether an applicant would be admitted to one public Midwestern physics graduate program. 

In addition, physics is a ``high consensus'' discipline, meaning most programs agree on what consists of a successful applicant \cite{posselt_inside_2016}. Therefore, despite many other components of the applications that affect whether an applicant will be admitted, we believe using the physics GRE score and undergraduate GPA provides a first-order overview of what admissions committees would use to admit applicants.

In order to ensure a reasonable number of applicants in each group to do meaningful analysis, we grouped applicants into bins based on their GPA and physics GRE score. We choose to use GPA bins 0.1 units in width and physics GRE bins 50 points in width. The GPA bins were selected to ensure that that GPAs with the same tenth digit were in a single bin. That is, 3.50 through 3.59 would be in a single bin. All GPAs were already reported on the 4.0 scale and physics GRE scores were reported using the standard 200-990 scale so we did not need to do any conversions. 

We then computed the fraction of applicants in each bin who were admitted to the program they applied. As we are interested in applicants ``standing out,'' we frame our results as whether applicants in a bin are admitted at a higher rate than the overall rate (all accepted applicants divided by all applicants). If applicants are admitted at a higher rate than the overall rate, it suggests that these applicants did in fact stand out to the admissions committee.

To take into account the size of the institution, we first used the AIP data to determine the national quartile each applicant's institution ranked in terms of all bachelor's degree recipients for each of the two years of data. Because not all institutions reported data in both years and the number of graduates could vary significantly between years, we conducted separate analyses first with the highest quartile an institution reached in the two years and second with the lowest quartile the program reached in the two years. For example, if an institution was ranked in the 3rd quartile the first year and the 4th quartile in the second year, our first analysis would use the 4th quartile and our second analysis would use the 3rd quartile. We then define the large programs as those in the 4th quartile and small programs as those in the 1st through 3rd quartiles. We address this choice in the discussion.

When using Barron's Selectivity Index to take into account the selectivity of the institution, we used Chetty et al.'s \cite{chetty_mobility_2017} five groupings (Ivy League +, Remaining most selective institutions, highly selective institutions, selective institutions, and non-selective institutions) as a guide. As there was a single applicant from a non-selective institution, selective and non-selective were grouped into a single category. Because we are interested in smaller, less known programs compared to larger, well-known programs, we took the selective and non-selective group to be our "less selective institution" group and institutions in the first three of Chetty et al.'s categories as our "most selective institutions". This corresponds to grouping institutions with a Barron's Index of 1 and 2 together as the "most selective institutions" and all other values together as the "less selective institutions".

To understand how high physics GRE scores might help applicants identifying as part of a group currently underrepresented in physics, we compared women's admission probability to men's admission probability and applicants of color's admission probability to applicants not of color's admission probability. While it should be noted that gender is not binary \cite{blue_gender_2018}, the data the admissions committee recorded is only in terms of the male and female binary and hence, we cannot comment on how high physics GRE scores may impact applicants identifying as other genders.

\begin{table*}[]
\caption{Counts of applicants by gender and race who provided both GPAs and physics GRE scores}
\label{demotable}
\begin{tabular}{lllllllll}
\cline{1-9}
      & \multicolumn{7}{c}{Race}                                  &       \\
Gender & Asian & Black & Latinx & Multi & Native & White & Unreported & Total \\ \hline
Men    & 247   & 49    & 99     & 166   & 4      & 1410  & 112     & 2087  \\
Women  & 56    & 2     & 19     & 26    & 0      & 308   & 28      & 439   \\
Unreported & 1 & 0 & 0 & 1 & 0 & 5 & 4 & 11 \\
Total & 304 & 51 & 118 & 193 & 4 & 1723 & 144 & 2537 \\  \cline{1-9} \cline{1-9}
\end{tabular}
\end{table*}

Further, given the limited number of applicants identifying as part of a racial group underrepresented in physics, we combined all applicants identifying as Black, Latinx, Multiracial, or Native into a single category, which we will refer to as B/L/M/N following the recommendation of Williams \cite{williams_underrepresented_2020}. We acknowledge that this may obscure important distinctions between groups, as Teranishi \cite{teranishi_race_2007} and Williams suggest. We also acknowledge applicants identifying as a marginalized gender and race may face additional barriers and hence could stand out differently than an applicant identifying as either a marginalized gender or race. However, there are less than 50 applicants (~2\% of the sample) identifying as a member of both a marginalized gender and marginalized race, limiting statistical power for analysis. Full demographics are shown in table \ref{demotable}. For information about how race and ethnicity categories were constructed and standardized, see Posselt et al. \cite{posselt_metrics_2019} who previously used the 2017 academic year application data from this study in their study.

\subsection{Mediation and Moderation Procedure}
Given that to some degree, both the physics GRE score and undergraduate GPA measure physics knowledge, we expect that these two measures will be correlated with each other. Therefore, we first tested whether the physics GRE and GPA have any mediating or moderating effects on each other when predicting admission. Because admissions status is a binary outcome variable, we need to use logistic regression for \cref{eqn1,eqn3,eqn4}.

When taking an applicant's GPA and physics GRE score into account, we first centered and scaled both variables so they both have means of zero and variances of 1. As we are treating GPA and physics GRE score as continuous, we can use linear regression for \cref{eqn2}.

To estimate the coefficients in \cref{eqn1,eqn2,eqn3,eqn4}, we generated 5000 bootstrap samples with replacement as was done in Hayes and Scharkow \cite{hayes_relative_2013}. For each trial, we computed the indirect effect $ab$. To get the estimate of each parameter, we took the average of the 5000 bootstraps. To get the lower end of the 95\% confidence interval, we used the value that corresponded to the 2.5th percentile of the values generated by the bootstrap. Likewise, to get the upper end of the 95\% confidence interval, we used the value that corresponded to the 97.5th percentile.

For the institutional features, we treat institutional selectivity and institution size as binary input variables (most selective or less selective and large institution or smaller institution) and the applicant's physics GRE score and GPA as continuous mediating and moderating variables. We then computed the coefficient estimates using the same bootstrap process as above. 

For demographic features, we treat gender and race as binary variables. Again, we use B/L/M/N as one category for race and white and Asian as the other. We also treat the applicant's physics GRE score and GPA as continuous mediating and moderating variables. We then computed the coefficient estimates using the same bootstrap process as above. 

\section{Results\label{sec:Results}}

\subsection{Probability of admission results}
\begin{figure*}
 \includegraphics[width=0.9\linewidth]{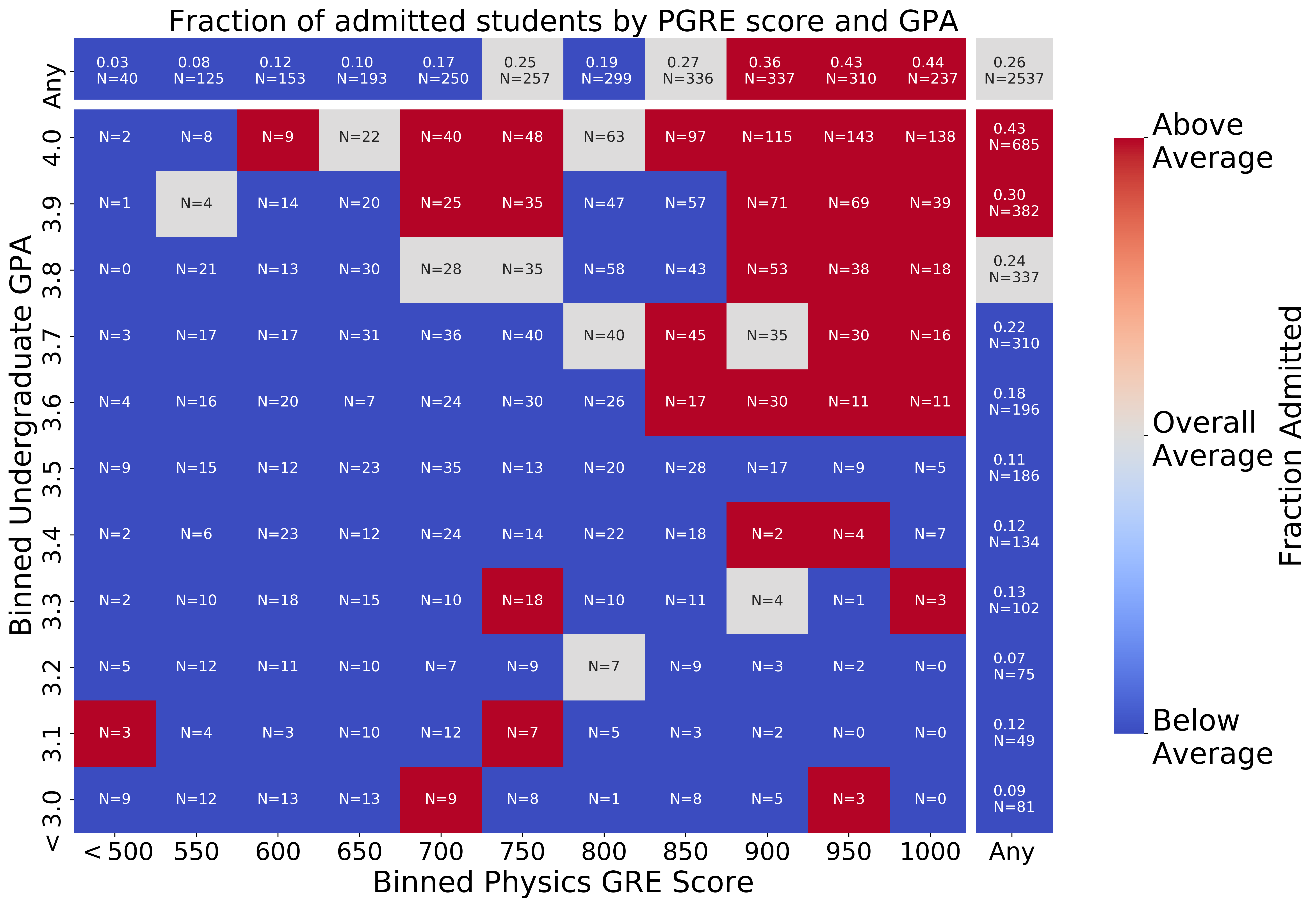}
 \caption{Fraction of applicants admitted by undergraduate GPA and physics GRE score. The number of students in each bin is also shown. `Any` corresponds to the corresponding row or column totals. The bin label corresponds to the upper bound of values in the bin exclusive with the exception of the 4.0 GPA bin which includes 4.0. Values are colored based on whether they are above, below, or equal to the overall admissions rate. Admissions rates within 10\% of the overall rate are colored the same as the overall rate. \label{gpa_pgre_fig}}
\end{figure*}

When comparing the GPAs and physics GRE scores of all students, we notice that most students who are admitted have both high GPAs and high physics GRE scores (Fig. \ref{gpa_pgre_fig}). Further, while a near perfect GPA or physics GRE score resulted in the highest chance of admission, having either a high GPA or high physics GRE and a modest score on the other seemed to still offer an admission fraction around the overall average. However, having a low GPA or low physics GRE and a modest score on the other is usually grounds for rejection. Overall admissions fractions for a given physics GRE score or GPA are shown in the top and right margins of Fig. \ref{gpa_pgre_fig} respectively.

\begin{figure}
 \includegraphics[width=1\linewidth]{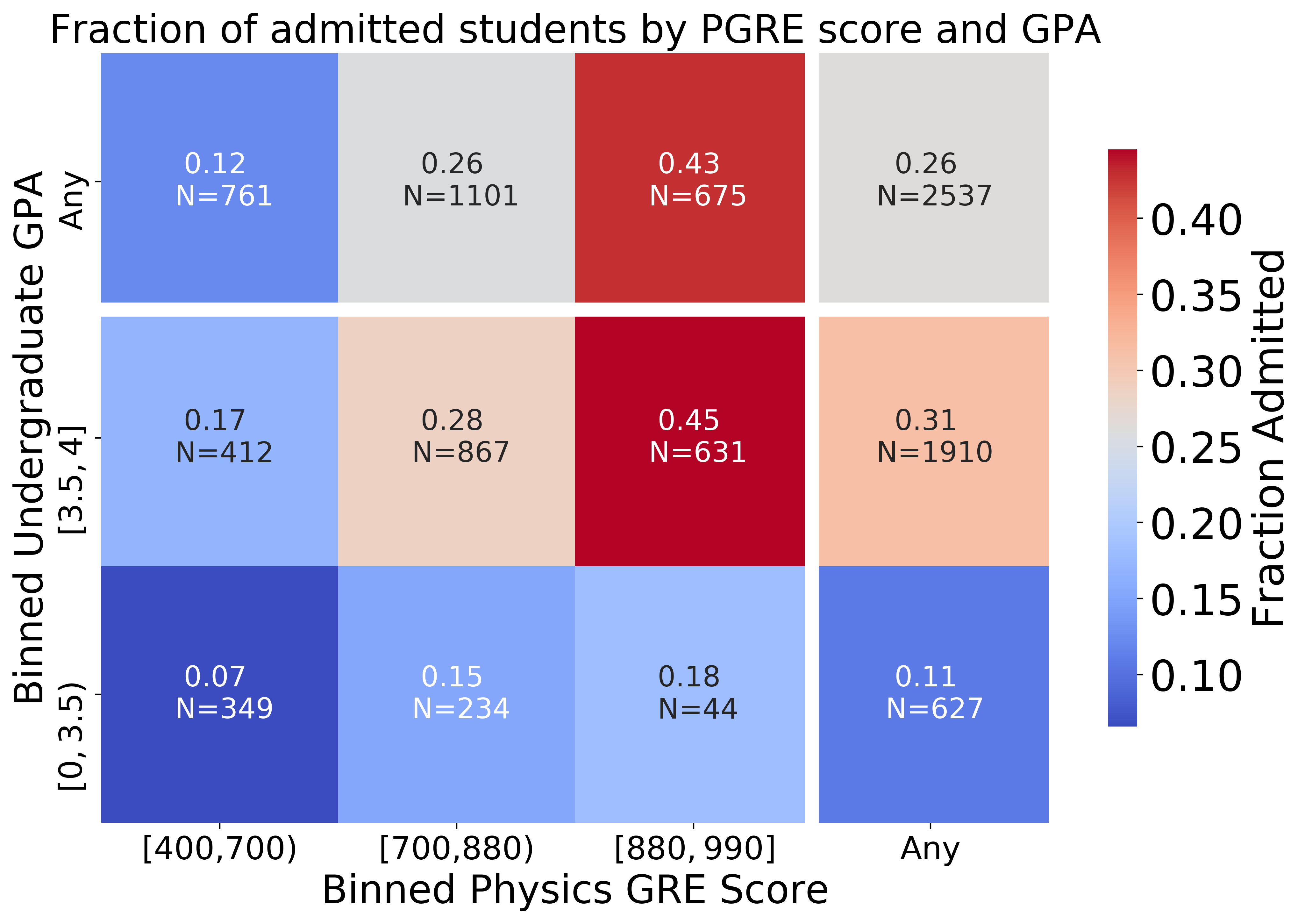}
 \caption{A condensed version of Fig. \ref{gpa_pgre_fig} showing the fraction of applicants admitted by undergraduate GPA and physics GRE score. \label{gpa_pgre_small}}
\end{figure}

When it comes to having a high physics GRE score despite a low GPA, we first note that only a small fraction of all applicants fall in this regime. Second, there appears to be no pattern in terms of higher than average fraction admitted for these applicants. Some combinations of low GPA and high physics GRE score result in a few applicants being admitted, and hence, an above average fraction of applicants being admitted, while other score combinations have no applicants being admitted, and hence, a below average change of admission. For example, having a GPA in the 3.3 bin and a physics GRE score in the 1000 bin resulted in an above average fraction admitted while having a GPA in the 3.4 bin and a physics GRE score in the 1000 bin did not result in an above average fraction admitted, despite the applicants having a higher GPA.

To further understand whether a high physics GRE score can highlight those with low GPA, we divided all students into either a high or low GPA and high or low physics GRE score bins, Fig. \ref{gpa_pgre_small}. Based on Fig. \ref{gpa_pgre_fig} in terms of admissions probabilities, a low GPA seems to be below a 3.5, while a high physics GRE score seems to be above 700. However, 700 is a common cutoff score which could explain why admissions probabilities increase after that score. Because hitting the minimum score might not catch the admission committee's eyes, we instead selected a higher score of 880 which represents the 80th percentile.

From Fig. \ref{gpa_pgre_small}, we notice two things. First, among applicants in the low GPA bin, less than half (44\%) even make it above the typical cutoff score of 700 and less than 10\% of those applicants with low GPAs score 880 or higher. These represent approximately 11\% and 2\% of all applicants respectively. Comparing the fraction of admitted applicants in each bin, applicants with high physics GRE scores and low GPAs are admitted at nearly the same rate as applicants with high GPA and low physics GRE scores.

Second, we notice that 16\% of all applicants score in the high GPA but low physics GRE score bin. That is, more applicants could be penalized for having a low physics GRE score despite a high GPA than could benefit from a high physics GRE score despite a low GPA.

\begin{figure*}
 \includegraphics[width=0.9\textwidth]{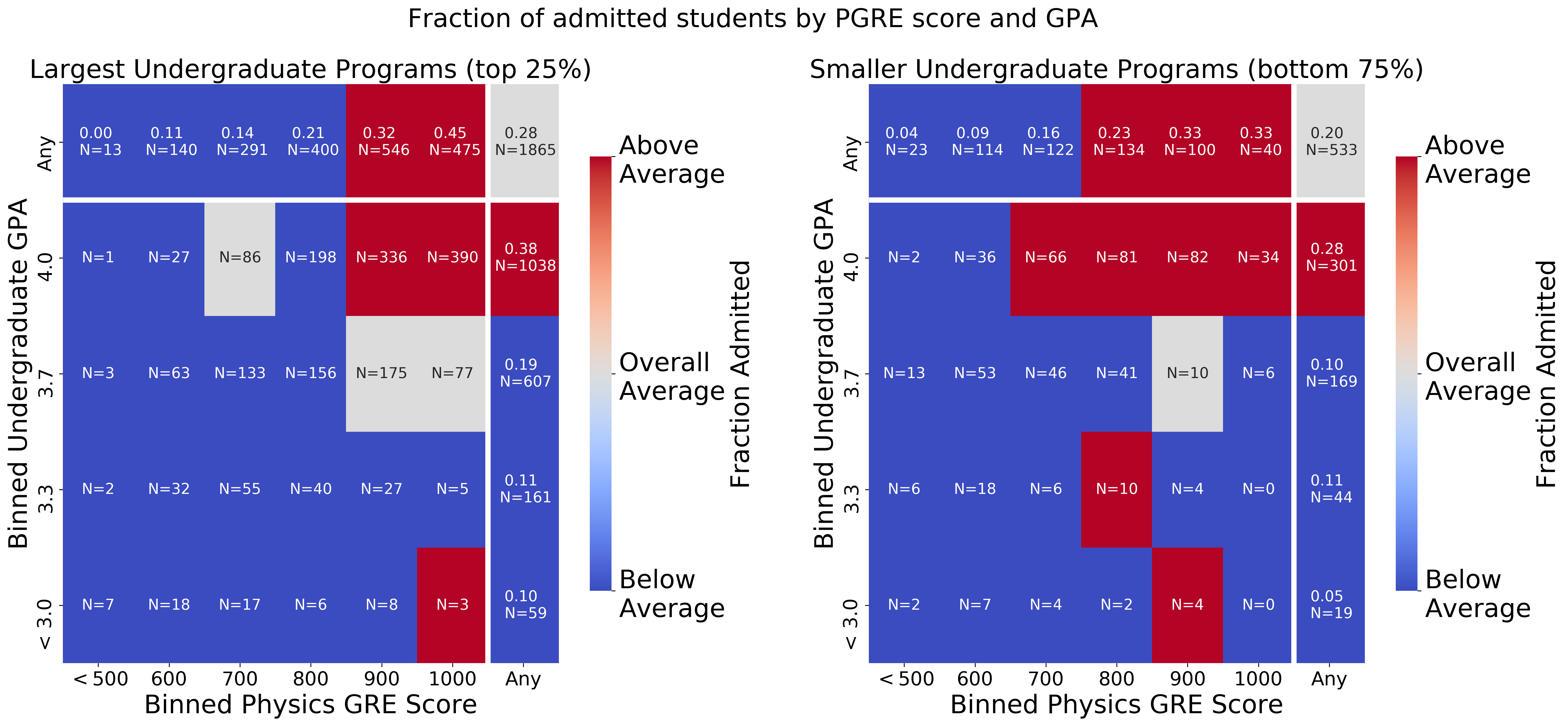}
 \caption{Fraction of applicants admitted by undergraduate GPA and physics GRE score and split by large or small undergraduate university\label{bach_fig}}
\end{figure*}

\begin{figure*}
 \includegraphics[width=0.9\textwidth]{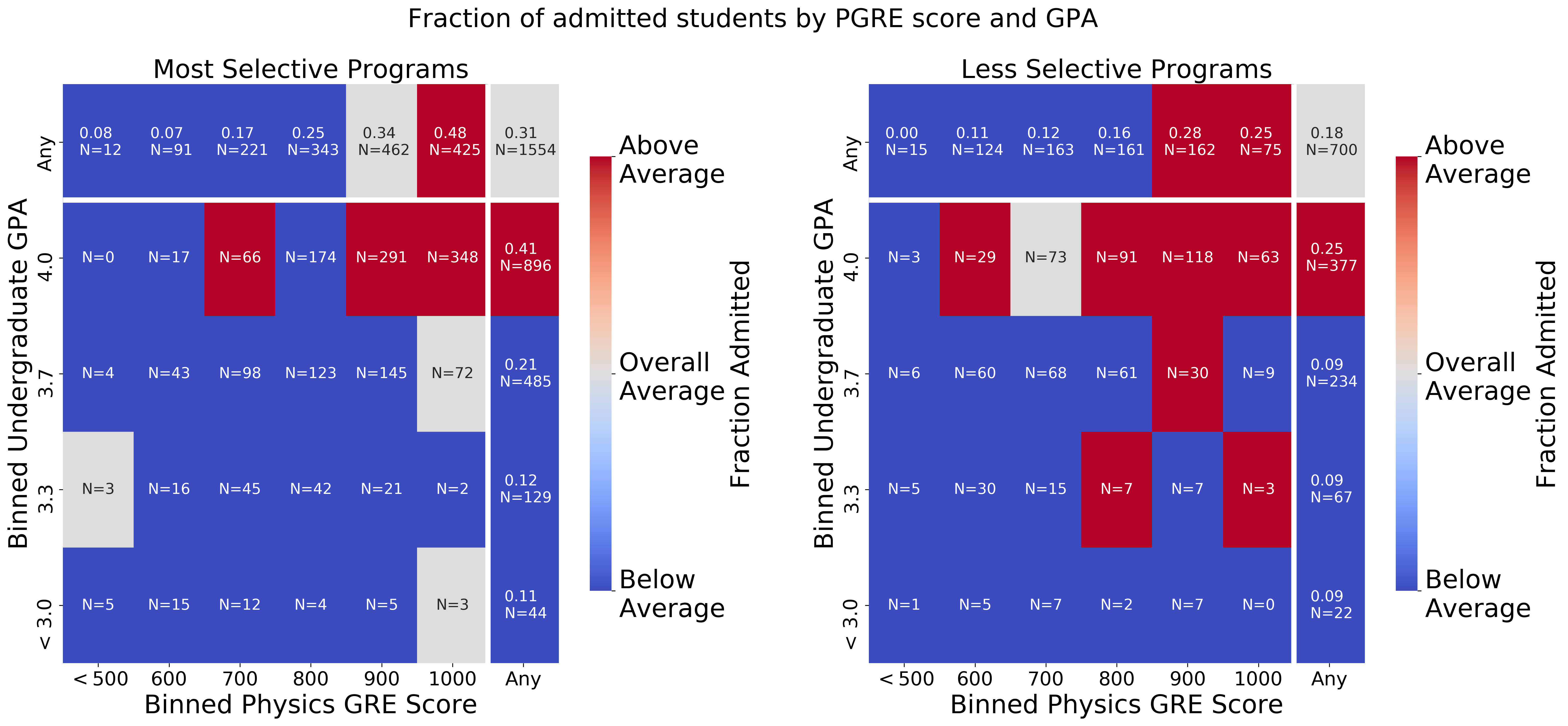}
 \caption{Fraction of applicants admitted by undergraduate GPA and physics GRE score and split by selective or non-selective undergraduate university.\label{fig_barron}}
\end{figure*}

\begin{table}[]
\caption{Distribution of applicants scoring in each Physics GRE range by size of institution. ETS only publishes overall score distributions and hence, we cannot report national scores from only domestic students.}
\label{gretable}
\begin{tabular}{llllll}
\hline
Score         & \begin{tabular}[c]{@{}l@{}}Large\\ Schools\end{tabular} & \begin{tabular}[c]{@{}l@{}}Small\\ Schools\end{tabular} & \begin{tabular}[c]{@{}l@{}}Selective\\ Schools\end{tabular} & \begin{tabular}[c]{@{}l@{}}Non-selective\\ Schools\end{tabular} & National \\ \hline
(400,500{]}   & 0.7\%  & 4.3\% & 0.8\% & 2.1\%  & 9\%  \\
{[}500,600)   & 7.5\%  & 21.4\% & 5.9\%  & 17.7\%  & 19\%  \\
{[}600,700)   & 15.6\% & 22.9\% & 14.2\% & 23.3\%   & 20\%  \\
{[}700,800)   & 21.5\%  & 25.1\% & 22.1\% & 23.0\%  & 19\%   \\
{[}800,900)   & 29.3\%  & 18.8\% & 29.7\%   & 23.1\% & 16\%   \\
{[}900,990{]} & 25.5\%   & 7.5\% & 27.3\%  & 10.7\% & 17\%   \\ \hline \hline
\end{tabular}
\end{table}

When taking the size of the applicant's undergraduate program into account, (large or small), using either the highest or lowest quartile of bachelor's graduates over the two year period did not substantially change the results. Therefore, we only present results from the highest quartile reached, which are shown in Fig. \ref{bach_fig}. Due to the much smaller number of applicants per bin, we reduce the number of GPA and physics GRE bins. We use bins of 3.0 or less, which corresponds to a B or lower, 3.0 to up 3.3, a B+, 3.3 up to 3.7, an A-, and 3.7 up to 4, an A under the standard 4.0 scale.

Overall, by looking at the bin in the `Any' row and `Any' column of Fig. \ref{bach_fig} and Fig. \ref{fig_barron}, we see that applicants from largest undergraduate programs are nearly 40\% more likely to be admitted (0.28 to 0.20) while applicants from selective institutions are nearly 70\% more likely to be admitted (0.31 to 0.18). Looking at the individual admission fractions, there does not appear to be any advantage to applicants graduating from smaller institutions or less selective institutions. The physics GRE scores and GPAs where applicants are admitted at higher than average rates are the nearly same for large and small programs and selective and non-selective programs. Unsurprisingly, these tend to be higher physics GRE scores and higher GPAs. Outside of a few bins with a small number of applicants, no combination of low GPA (B+ or less) and high physics GRE score resulted in above average admission fraction.

For the highest physics GRE scores, 900 and above, applicants from the largest or most selective universities seem to be admitted at a higher rate and a higher fraction of applicants from large or selective universities achieve these high scores compared to applicants from smaller universities. The fraction of applicants from both large universities, small universities, selective universities, and non-selective universities, as well as nationally, achieving each score is shown in table \ref{gretable}. Thus, it appears that even if higher scores did help applicants stand out, applicants from smaller and less selective schools most in need of standing out are less likely to achieve those scores in the first place.

\begin{figure*}
 \includegraphics[width=0.9\textwidth]{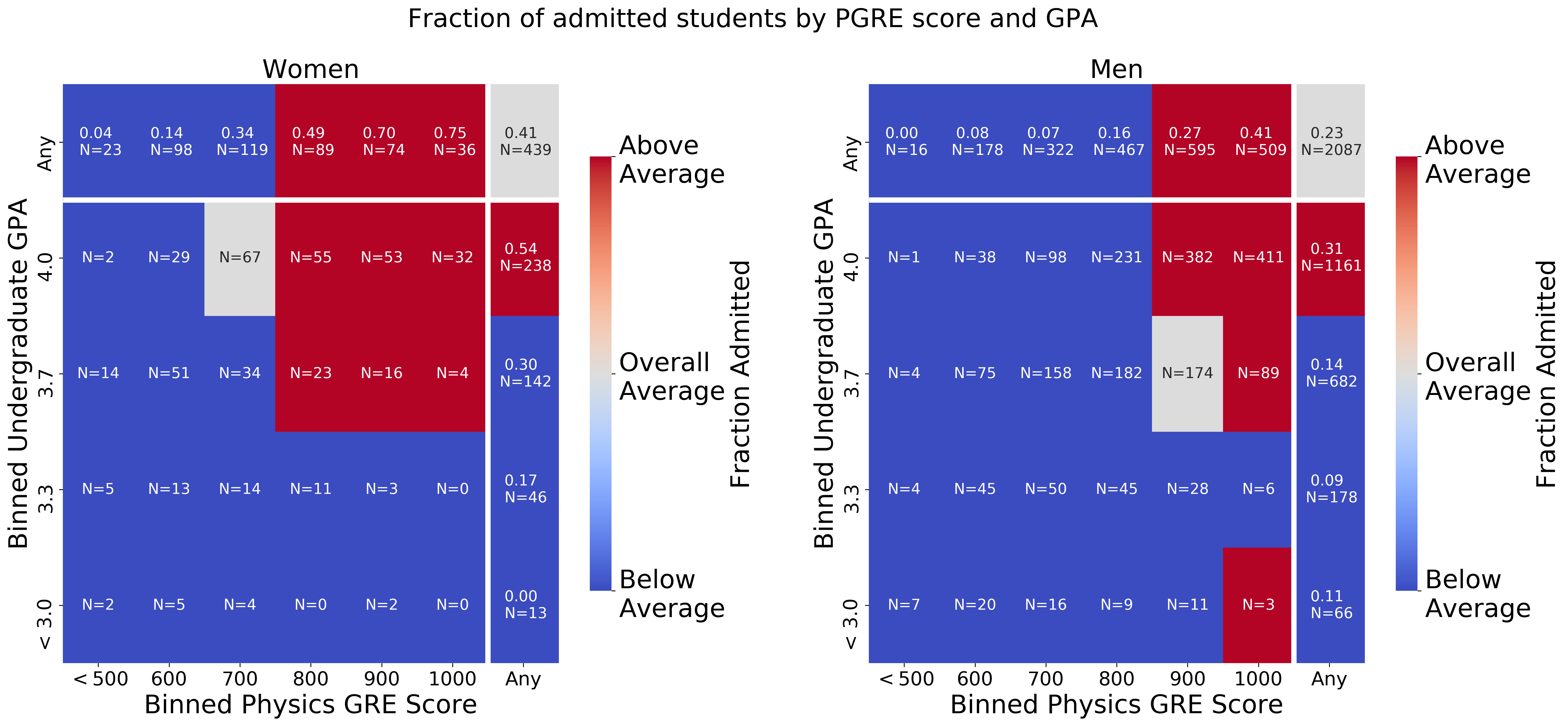}
 \caption{Fraction of applicants admitted by undergraduate GPA and physics GRE score and split by the applicant's gender.\label{gender_fig}}
\end{figure*}

\begin{figure*}
 \includegraphics[width=0.9\textwidth]{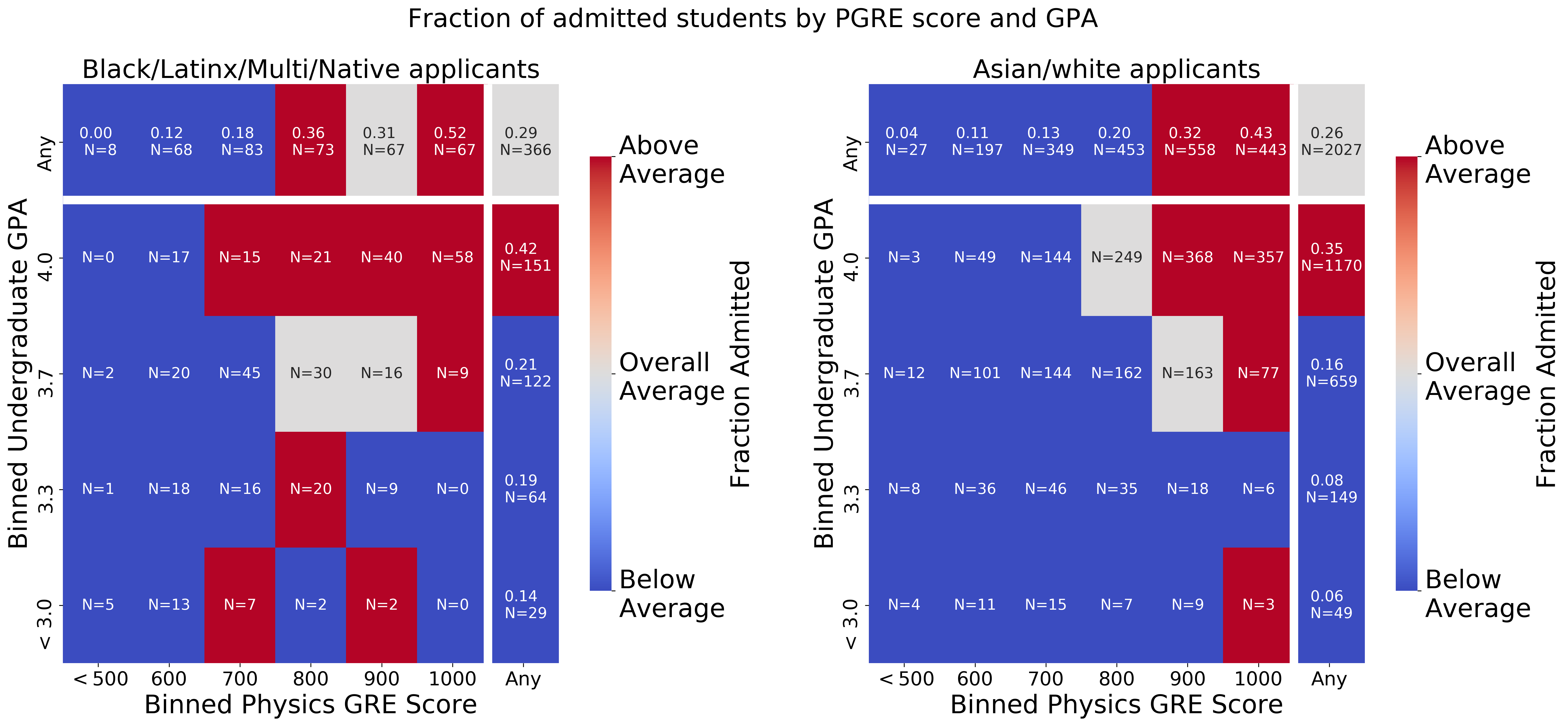}
 \caption{Fraction of applicants admitted by undergraduate GPA and physics GRE score and split by the applicant's race.\label{race_fig}}
\end{figure*}

Finally, the results from grouping by gender and race are shown in Fig. \ref{gender_fig} and Fig. \ref{race_fig}. Interestingly, we find that for most physics GRE scores, women are admitted at higher rates than men of equal score are. Likewise, we find that applicants identifying as Black, Latinx, Multi-racial, or Native are admitted at higher or the same rates at applicants identifying as white or Asian for similar physics GRE scores. In addition, the same trend seems to hold for GPA as well. However, a high physics GRE score does not seem to help women with a low GPA. For B/L/M/N applicants, there appears to be a few places where applicants may stand out (such as the 800 physics GRE bin and 3.3 GPA bin). If these applicants were standing out due to their physics GRE score though, we would expect that pattern to continue for higher physics GRE scores but the same GPA. This does not appear to be the case, suggesting these applicants stood out for a reason other than their physics GRE scores. We address this in our discussion.

Because these applicants may have stood out for reasons other than their physics GRE score, we do not discuss any interactions between gender and race and selectivity and institution size. For completeness, plots showing these interactions are included in the supplementary material.

\subsection{Mediation and moderation results}
\subsubsection{Physics GRE and GPA}
\begin{figure}
 \includegraphics[width=1\linewidth]{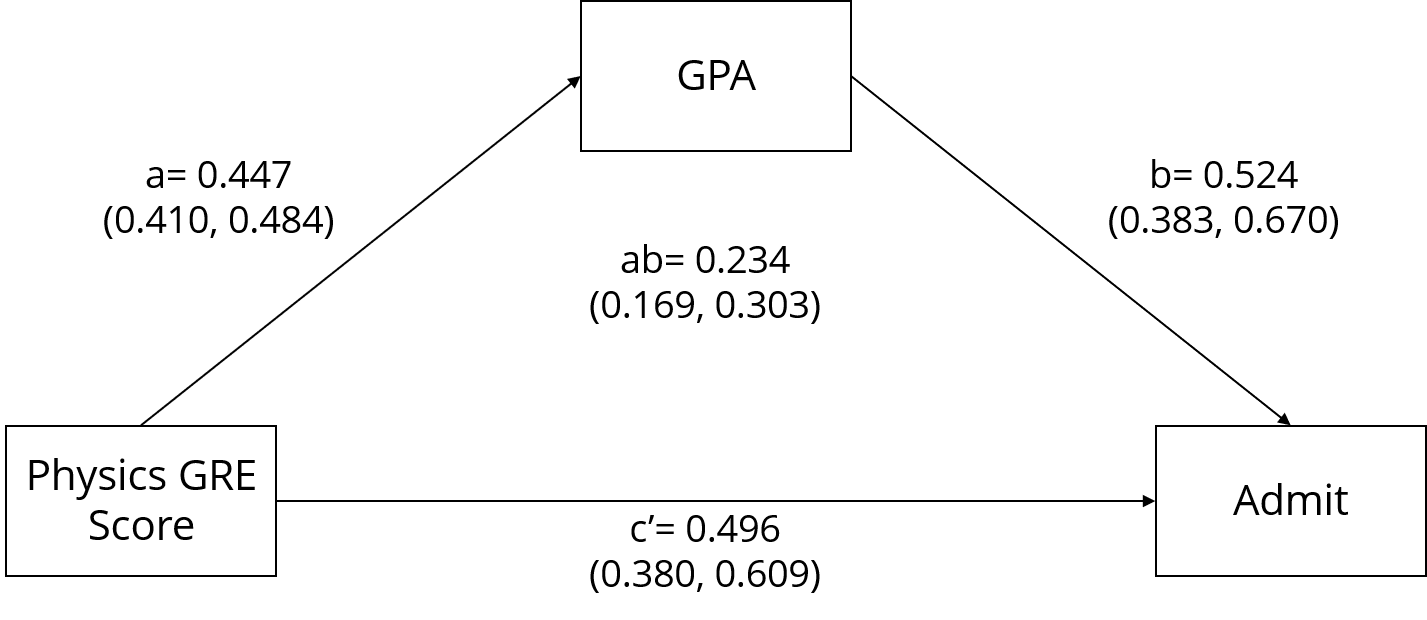}
 \caption{Visual representation of the bootstrapped coefficients in \cref{eqn1,eqn2,eqn3}. We do find evidence of GPA mediating the relationship between physics GRE score and admission status.\label{med_results}}
\end{figure}

Whether we pick the physics GRE score or GPA as the mediating or moderating variable does not change the results, so because our previous work showed that the physics GRE has more predictive power than the applicant's GPA for admission \cite{young_using_2020}, we only present the results when the independent variable is the physics GRE score. A visual representation of our mediation results with the physics GRE score and GPA is shown in Fig. \ref{med_results}. We find that all coefficients are statistically different from zero.

From Fig. \ref{med_results}, we see that an applicant's physics GRE score and GPA have about the same effect on whether the applicant is admitted. Given that applicants who had either a high physics GRE score or a high GPA had about the same chance of being admitted, this is not a surprising result.

Second, we find that the indirect effect is not zero, meaning that there is partial mediation. That is, whether an applicant is admitted depends on their physics GRE score and their GPA. In terms of the amount of moderation, we find that the indirect effect accounts for nearly 33\% of the total effect.

Finally, doing moderation analysis, we find that $b_2=0.024 \; (-0.114,0.154)$. As zero is included in the confidence interval, we do not find evidence that the physics GRE score moderates the relationship between an applicant's GPA and whether they are admitted. That is, relationship between an applicant's GPA and whether they are admitted is not influenced by their physics GRE score.

\subsubsection{Institutional features}
\begin{figure*}
 \includegraphics[width=1\linewidth]{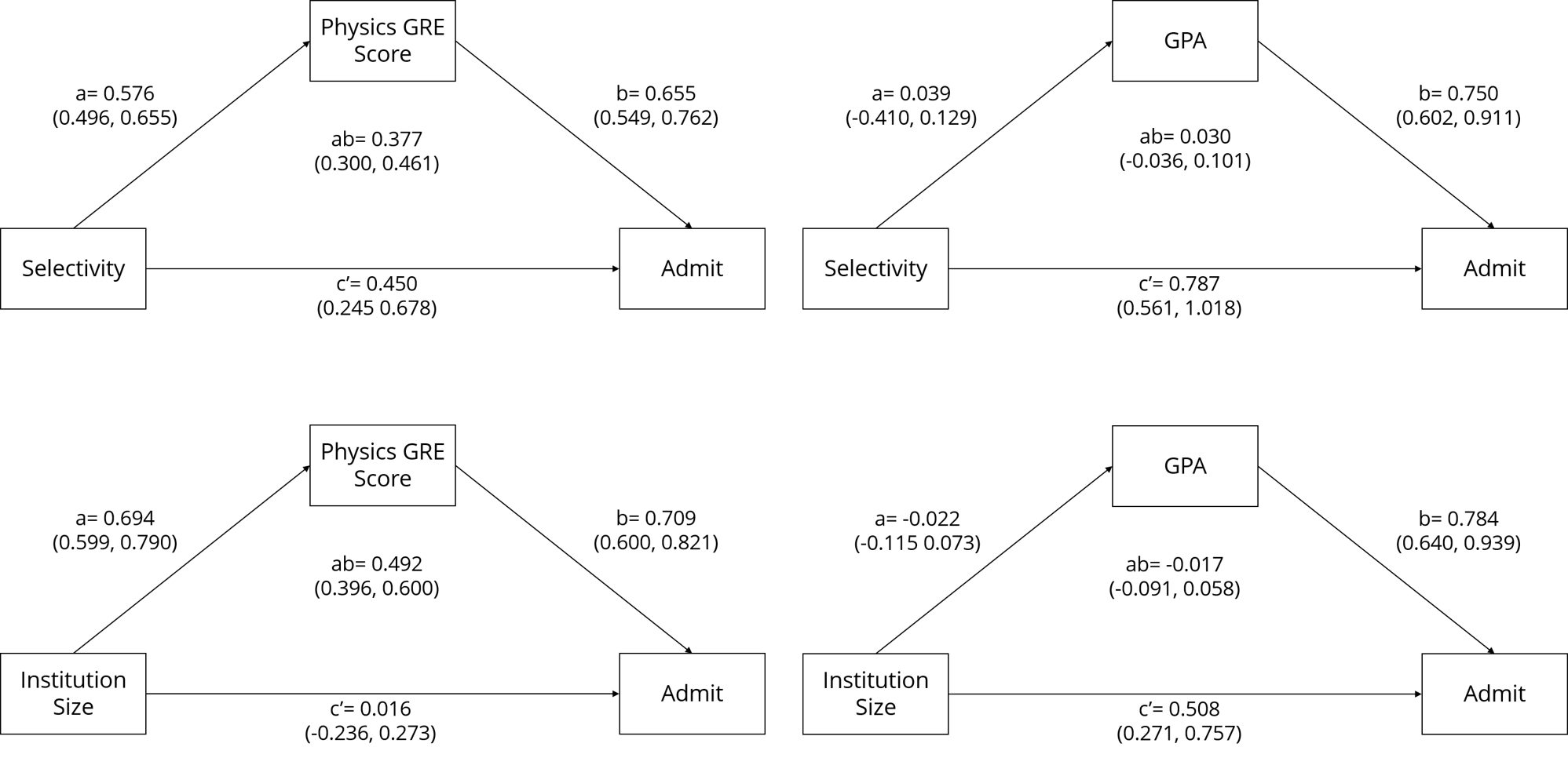}
 \caption{Visual representation of the bootstrapped coefficients in \cref{eqn1,eqn2,eqn3}. We do find evidence of the physics GRE score mediating institution size and selectivity but do not find evidence of GPA mediating institution size or selectivity. \label{med_barrons_size}}
\end{figure*}

A visual depiction of our results is shown in Fig. \ref{med_barrons_size}. We find that the applicant's physics GRE score partially mediates the relationship between the selectivity of their undergraduate institution and whether they were admitted and fully mediates the relationship between their institution's size and whether they were admitted. The fractions of mediation due to the indirect effects were $\frac{ab}{ab+c'}= 0.456$ and $0.969$ respectively.

In contrast, the applicant's GPA was not found to be a significant mediator in either case (zero was contained in the indirect effects' 95\% confidence intervals).

When looking at the results of the moderation analysis when the physics GRE is the mediating variable, we find that neither $b_2$ value is statistically different from zero ($b_{2,selectivity}=0.235 \; (-0.015,0.476)$ and $b_{2,institution_size}=0.104 \; (-0.161,0.3.63)$). However, in the case of institutional selectivity moderating physics GRE and admit, the result is marginally significant. Indeed, zero is not contained in the 93\% confidence interval for $b_{2,selectivity}$.

\subsubsection{Demographic features}
\begin{figure*}
 \includegraphics[width=1\linewidth]{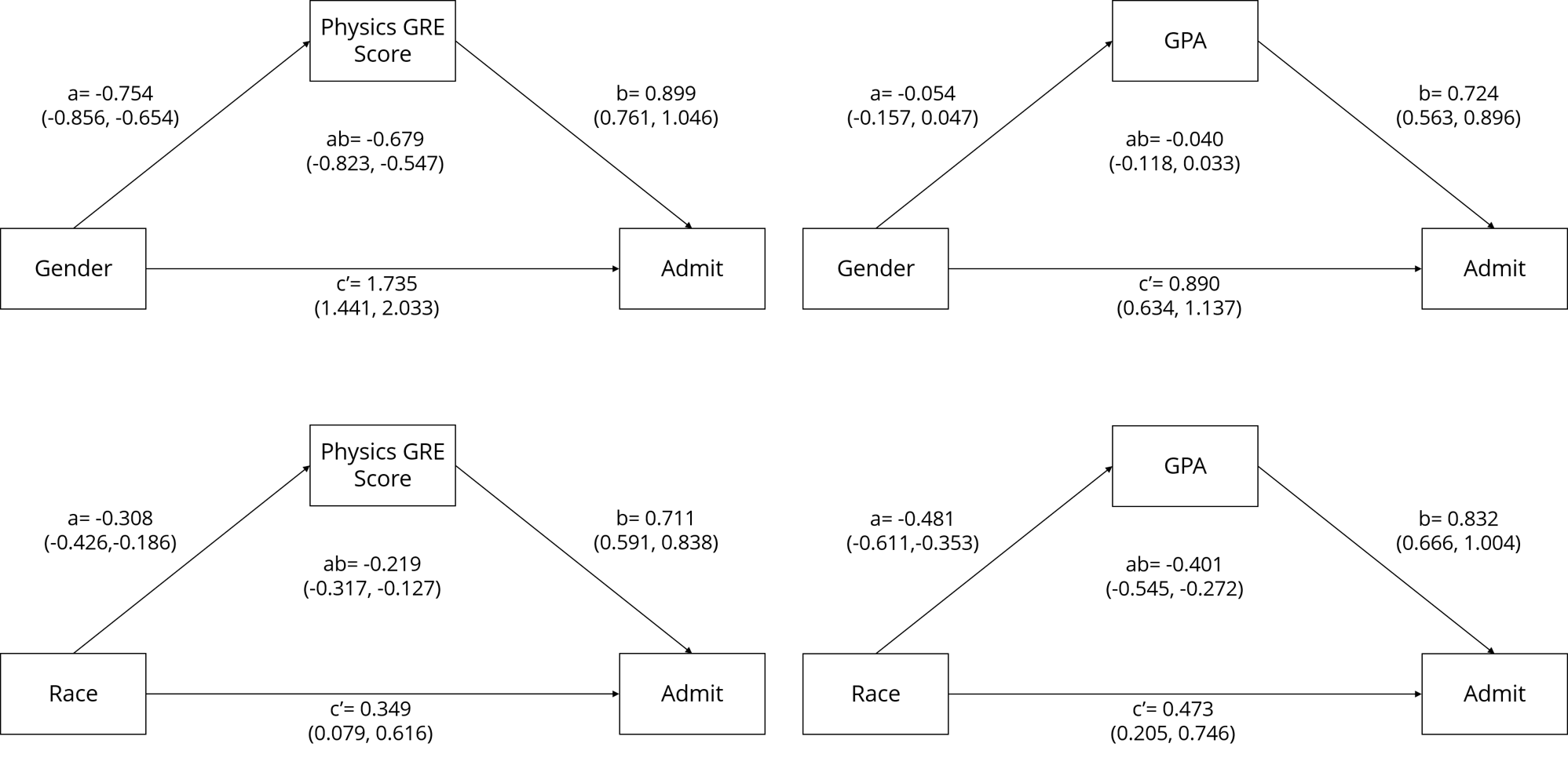}
 \caption{Visual representation of the bootstrapped coefficients in \cref{eqn1,eqn2,eqn3}. We do find evidence of the physics GRE score mediating gender and race and GPA mediating race but do not find evidence of GPA mediating gender. \label{med_gender_race}}
\end{figure*}

Our results are shown visually in Fig. \ref{med_gender_race}. Because we chose woman to be "1" and B/L/M/N to be "1" in our logistic regression equation, some of the coefficients are negative. For example, the negative $a$ coefficient for gender and physics GRE score means that women score lower on the physics GRE than men do. Because the sign depends on our choice of which category should be "1" and are in that sense arbitrary, we use the absolute values of $c'$ and $ab$ to calculate the fraction of mediation.

We find that the applicant's physics GRE score partially mediates the relationship between both gender and admission and race and admission. The fractions of mediation for gender and admission is $\frac{|ab|}{|ab|+|c'|}=0.281$ and for race and admissions is $\frac{|ab|}{|ab|+|c'|}=0.386$.

We also find that GPA partially mediates the relationship between race and admission but not gender and admission. The fraction of mediation for race and admission due to GPA is $\frac{|ab|}{|ab|+|c'|}=0.458$ respectively.

When investigating whether any moderation effects exist, we do not find that to be the case. That is, we find that none of the $b_2$ values are statistically different from zero. Specifically,
\begin{itemize}
    \item $b_{2,pGRE,gender}=0.192 \; (-0.090,0.483)$, 
    \item $b_{2,GPA,gender}=0.241 \; (-0.072,0.578)$,
    \item $b_{2,pGRE,race}=-0.011 \; (-0.266,0.259)$,
    \item $b_{2,GPA,race}=-0.209 \; (-0.517,0.126)$.
\end{itemize}

All results and interpretations from the mediation and moderation analyses are summarized in table \ref{tab:medmod_table}.

\begin{table}[]
\caption{Summary of the mediating and moderation results. * signifies partial mediation is present, ** signifies full mediation is present, $\dagger$ signifies moderation is present. However no moderation effects were found.}
\label{tab:medmod_table}
\begin{tabular}{llll}
\hline
Independent & Mediating & Indirect effect & Moderating effect \\ \hline
Physics GRE & GPA & 0.234* & 0.024 \\
Selectivity & Physics GRE & 0.377* & 0.235 \\
Selectivity & GPA & 0.030 & 0.132 \\
Institution size & Physics GRE & 0.492** & 0.104 \\
Institution size & GPA & -0.017 & -0.109 \\
Gender & Physics GRE & -0.679* & 0.193 \\
Gender & GPA & -0.040 & 0.241 \\
Race & Physics GRE & -0.219* & -0.011 \\
Race & GPA & -0.401* & -0.201 \\ \hline
\end{tabular}
\end{table}

\section{Discussion}\label{sec:Discussion}
Here, we address each of our research questions and possible limitations or confounding factors.

\subsection{Research Questions}
\emph{How does an applicant's physics GRE score and undergraduate GPA affect their probability of admission?} We find that scoring highly on the physics GRE and having a high GPA results in the highest chance of admission (Fig. \ref{gpa_pgre_small}). Likewise, having a low physics GRE score and low GPA results in the lowest chance of admission. If either the applicant's physics GRE score or GPA is high while the other is not, the chance of admission is approximately equal, regardless of which one is high. 

However, the number of applicants with high GPAs but low physics GRE scores is 9 times as large as applicants with low GPAs and high physics GRE scores (i.e. scoring above the 80th percentile; Fig. \ref{gpa_pgre_small}). Even if we consider meeting the minimum cutoff score as a high physics GRE score, the number of applicants who have high GPAs but low physics GRE scores is 1.5 times greater than the number of applicants with low GPAs but high physics GRE scores. Thus, many more high GPA applicants could be penalized by the physics GRE than low GPA applicants could stand out or benefit with a high physics GRE score.

Finally, we note that for low-GPA applicants with high physics GRE scores, they are all essentially admitted at the same rate, regardless of whether they scored in the 700-870 range or the 880-990 range. If these applicants were standing out, we would expect low GPA applicants scoring above 880 to be admitted at a much higher rate than low GPA applicants scoring between 700 and 870. Thus, it is hard to determine if these applicants actually stood out to the committee or if they simply met the minimum physics GRE score needed for the committee to review the rest of the application.

\emph{How are these probabilities of admission affected by an applicant's undergraduate institution, gender, and race?} First, we find that for most physics GRE scores, applicants from larger and smaller institutions are admitted at similar rates (Fig. \ref{bach_fig}). However, for the highest scores (above 900), applicants from larger universities are admitted at higher rates. Interestingly, for applicants from smaller programs scoring above 900 does not appear to provide any additional benefit in terms of the fraction of applicants admitted compared to scoring between 800 and 900.

In contrast, applicants from less selective institutions are less likely to be admitted than applicants from more selective institutions for all physics GRE scores above the common cutoff score (Fig. \ref{fig_barron}). That is, the physics GRE does not seem to counteract any potential biases from admissions committees toward applicants from less selective institutions.

Overall, attending a large or selective institution and scoring highly on the physics GRE does result in a higher chance of admission than scoring highly on the physics GRE and attending a smaller or less selective institution.

It is important to note that there might be selection bias in our data because test-takers with high scores from smaller universities might not chose to apply to these schools. However, this seems unlikely because 1) these programs are highly regarded and hence, these would not be "safety schools" to high scoring applicants (as indicated by many high scoring applicants from large programs applying here) 2) while there is research suggesting students with low physics GRE scores might view their scores as barriers to applying \cite{cochran_identifying_2018}, to our knowledge, there is no evidence that students with high scores do not apply to physics graduate programs. Given that students with low test scores might not apply, it is expected that our data is not representative of test-takers on the lower end of scores (as shown in table \ref{gretable}). 

When looking at the demographic variables, we find that women are admitted at higher rates than men with similar scores (Fig. \ref{gender_fig}) and B/L/M/N applicants are also admitted at higher rates than white or Asian applicants (Fig. \ref{race_fig}. As prior work has shown \cite{miller_test_2014}, women and B/L/M/N test-takers tend to score lower than white men on the physics GRE and hence, scoring highly could cause these applicants to stand out to admissions committees.

\emph{How might the above relationships be accounted for through mediating and moderating relationships?}
Our mediation and moderation analysis further supports the results found through the probability of admissions procedure. 

We find that the physics GRE score and GPA have similar regression coefficients when modeling admission, suggesting they have similar effects (Fig. \ref{med_results}). In addition, we did not find any evidence of moderation. That means the relationship between GPA and admission is not different due to the applicant's physics GRE score. If a high physics GRE score did help a low-GPA applicant stand out, we would expect to see a moderation effect.

Combining the results of probability of admission analysis and the mediating and moderation analysis, we find that there is no interaction between an applicant's physics GRE score and their GPA when it comes to admission probability. An applicant with a low GPA cannot simply overcome that low GPA by scoring highly on the physics GRE.

When we performed mediation analysis on the institutional factors, we found that the relation between institutional selectivity and admission was partially mediated by the applicant's physics GRE score and the relation between institutional size and admission was fully mediated by the applicant's physics GRE score (Fig. \ref{med_barrons_size}). Neither of these relationships was mediated by the applicant's GPA however.

The results of the mediation analysis show that physics GRE scores seem to explain some of the differences in admission probability based on the applicant's undergraduate institution. Therefore an applicant from a smaller or less selective institution may be able to stand out by scoring highly on the physics GRE. However, looking at the fraction of applicants by physics GRE scores, especially the highest scores, suggests that is not what happens in practice.

Our moderation analysis found a marginally significant difference in the relation between an applicant's physics GRE score and admission status by institution selectivity, which warrant further analysis. If future work does find a moderation effect from an applicant's physics GRE score on the relation between their institutional selectivity and admission status, it would challenge the claim that the physics GRE offers a level playing field for all applicants.

In terms of gender and race, we do find some mediating relationship, but no moderation relationships (Fig. \ref{med_gender_race}). We find that the physics GRE partially mediates the relationship between gender and admission and the relationship between race and admission. We also find GPA partially mediates the relationship between race and admission. That is, some of the differences in admission rates between men and women can be explained by the differences in their physics GRE scores and some of the differences in admission rates between B/L/M/N applicants and non-B/L/M/N applicants can be explained by differences in their physics GRE scores and GPAs.

These results then suggest that a female or B/L/M/N applicant may be able to stand out by doing well on the physics GRE. In practice, the probability of admission results do suggest that women and B/L/M/N applicants are admitted at higher rates than their male, white, or Asian peers are. However, as the five programs studied here were interested in increasing their diversity, our data does not allow us to disentangle "standing out" from highlighting. Therefore, our results should be interpreted with caution regarding any claims that the physics GRE may help applicants from groups underrepresented in physics stand out. 

It should also be noted that women and B/L/M/N applicants are less likely to reach these higher scores than their male, white, and Asian peers. In our data, 75\% of men and 72\% of white or Asian applicants scored above 700 compared to 45\% of women and 57\% of B/L/M/N applicants. Thus, even if the physics GRE does allow these applicants to stand out, any potential benefit must be weighed against known scoring discrepancies.

\subsection{Limitations and Researcher Decisions}{\label{sec:limitations}}

\emph{Data Biases}
As previously noted, applicants with lower physics GRE test scores may be less likely to apply, resulting an over-representation of high scoring applicants. In addition, the programs in this study are well-regarded programs and there is likely a secondary bias toward applicants with high GPAs and high physics GRE scores applying overall. As a result, the results may not generalize to graduate programs whose applicants tend to have lower GPAs or low physics GRE scores.

In addition, it is possible that an applicant could be represented multiple times in the data set, as an applicant could have applied to more than one of the five universities in this study. However, each applicant applies to each program independently and thus, we can treat them as separate events for the admissions probabilities. On the other hand, results based on distributions such as table \ref{gretable} and Fig. \ref{rain_inst} and Fig. \ref{rain_demo} would be affected by the duplicates. When we compare the distributions both with and without possible duplicates, Kolmogorov-Smirnov tests \cite{massey_jr_kolmogorov-smirnov_1951} suggest the distributions are not significantly different. Therefore, because we cannot actually determine which applicants are duplicates and excluding possible duplicates does not change our results, we did not remove possible duplicates.

\emph{Our choice of low GPA and high physics GRE} While percentiles are available for the physics GRE, a ``high score'' is left to interpretation. Even among admissions committees, individual members may have different ideas of what a high score is. In our work, we have taken the common cutoff score of 700 as the minimum possible high score \cite{miller_typical_2019}. Even around this minimum score, the number of applicants with low GPAs who could benefit from scoring highly on the physics GRE is less than the number of high GPA applicants who could be penalized by having a score below the cutoff. 

We find that the number of low GPA applicants who could benefit from a high physics GRE score is approximately equal to the number of high GPA applicants who could be penalized by a low score when the high score cutoff is 670, which is lower than the typical cutoff score and is around the 43rd percentile. Assuming a high score should be at least above the 50th percentile, our specific choice of a high score does not affect our result that more applicants could be penalized than could benefit.

The previous argument is also affected by what we consider a high GPA. We have chosen any GPAs less than 3.5 to be low based on the results shown in Fig. \ref{gpa_pgre_fig} where applicants with GPAs at or above 3.5 are nearly twice as likely to be admitted to as applicants with GPAs below 3.5. If we were to pick a lower threshold, there would be even fewer applicants in the low GPA-high physics GRE score group and more applicants in the high GPA-low physics GRE score group, meaning even more applicants would possibly be penalized rather than standout. If we instead picked a higher GPA such as 3.6, there would be more applicants who could potentially benefit, but even then, the number of applicants who could benefit is nearly equal to the number of applicants who could be penalized around a physics GRE score of 730, which is not a high physics GRE score (approximately 54th percentile) and does not significantly change our results. If we were to pick an even higher GPA cutoff, we could be hard-pressed to justify why anything other than an `A' GPA is considered a low GPA, especially because admissions committees seem to group applicants with GPAs between 3.5 and 3.6 more closely with applicants with GPAs between 3.7 and 3.8 than applicants with GPAs between 3.4 and 3.5 (based on the fraction of applicants admitted). 

Based on our data and the fact that some universities use 3.5 as the only separation between a 3.0 and 4.0, using 3.5 seems to represent the best option for separating high and low GPA students. Using any other choice either strengthens our claims or seems unrealistic to use as a cutoff.

\emph{Our choice of non-selective school} 
We choose to follow a modified version of Chetty et al.'s groupings of programs \cite{chetty_mobility_2017}. However, many large, state universities have a Barron's Selective Index of 3 and fall in Chetty et al.'s fourth group. For our analysis, we would have included these large, state institutions as part of the less selective programs. As we are concerned with whether the physics GRE helps applicants stand out, saying applicants from large, state universities (for example, the University of Colorado-Boulder, the University of Washington, and Michigan State University) may fall in the traditionally missed category may not be correct.

We reran the analysis with these large, state institutions as part of what we called the most selective programs. We find that the conclusions are then more aligned with the large vs small program results. Using this grouping, applicants from less selective programs are admitted at similar rates to applicants from more selective programs for most physics GRE scores. However, applicants from more selective institutions with physics GRE scores above 900 are still more likely to be admitted than applicants from less selective institutions with similar physics GRE scores. 

In terms of the mediating and moderation analysis, our results would be strengthened under this choice. The Physics GRE score fully mediates the relationship between selectivity and admission. In addition, selectivity moderates the relationship between the physics GRE and admission. That is, under this grouping, we do find that the relation between physics GRE score and admission depends on the applicant's undergraduate institution, with a high score from a selective institution carrying more "weight."

Thus, even though the details change, the overall conclusion are not weakened by changing our groupings. In fact, changing the groupings may strengthen our conclusions instead.

\emph{Our choice of a ``small'' school} We chose to small schools to be any university not in the top quartile of yearly bachelor's degrees awarded. We acknowledge that using quartiles is a somewhat arbitrary decision. However, when we used halves instead of quartiles to divide large and small school, our results were unchanged, both in terms of the probability of admission analysis and the mediation and moderation analysis. Using the bottom quartile as small schools and all other programs as the large school would not have yielded insightful results as less than 2\% of applicants would have attended a small school under this choice.


Of the possible physics specific measures, the number of bachelor's degrees seems most appropriate because programs with more graduates are more likely to be known by admissions committees simply because there are more students to apply from those programs. For example, the programs in the top quartile by number of bachelor's graduates produce nearly two-thirds of all physics bachelor's graduates \cite{nicholson_roster_2019,nicholson_roster_2018}. In addition, we assume that programs with strong physics reputations attract more students and hence, produce more graduates. While this is likely to be more true at the graduate level, not all physics programs offer graduate degrees and hence, using the number of PhDs awarded would not be useful. Thus, we believe the number of bachelor's graduates serves as a rough proxy for physics reputation.

\section{Future Work}\label{sec:FutureWork}
While the five universities included in this study were interested in increasing their diversity and reducing inequities in their program, their admissions processes still resembled the traditional metrics-based admissions model. Recently, many programs, including the ones studied here, have begun to employ holistic admissions, which looks at the overall application, taking into account non-cognitive competencies and contextualizes the accomplishments of the applicant in terms of the opportunities that were available to them \cite{miller_broadening_2018,kent_holistic_2016}. Often these holistic admissions use rubrics to weight the various components of each applicant (e.g. see \cite{rudolph_final_2020,dunlap_journey_2018}). Evidence from biomedical science graduate programs suggests that the GRE can even be included in holistic admissions without reproducing its known gender and racial biases \cite{wilson_model_2019}. Further, their two-tiered approach to holistic admissions did not significantly increase the workload of admission committee members. These findings could persuade faculty reluctant to remove GRE due to its ease and supposed ability to measure some innate quality to try holistic admissions. Whether these results would hold for decentralized admissions as is typical in physics and for the physics GRE though are still open questions.

Our future work will then examine how our results may be affected when a department uses holistic admissions. In theory, we should no longer see the discrepancies between admitted applicants from large and small programs and more selective and less selective universities. In addition, the sample rubric developed by the Inclusive Graduate Education Network (as shown in \cite{rudolph_final_2020}) suggests ranking applicants by high, medium, or low on each part of their applicant. Therefore, we would expect to see a flatter distribution of admission fractions based on physics GRE scores because for example, all scores within the `high' range should be treated equally in the admissions process. Our future work will determine if this is indeed the case.

\section{Conclusion and Implications}\label{sec:Conclusion}
Our work suggests that scoring highly on the physics GRE does not help applicants from small or less selective schools or applicants with a low GPA "stand out". Indeed, having a high physics GRE and low GPA is no better than having a low physics GRE score and high GPA in terms of the fraction of applicants admitted. Similarly, for average physics GRE scores, the selectivity or size of the applicant's institution does not offer any advantage. For the highest scores though, attending a smaller or less selective institution does appear to result in an admissions penalty.

We do find that women and B/L/M/N applicants do have higher rates of admission based on physics GRE scores. However, given that the departments included in this study were actively trying to improve the diversity of their graduate student population \cite{posselt_metrics_2019}, we are unable to attribute that standing out to the physics GRE.

In response to the ETS's claim that the physics GRE can help applicants stand out from other applicants, we do not find evidence to support that claim. In fact, our results suggest the opposite: the physics GRE may penalize applicants due to a low score rather than help applicants due to a high score. 

As Small points out, facts and data do not unambiguously prescribe a course of action \cite{small_range_2017} and as other have noted, making such courses of action require a framework of assumptions and commitments \cite{searle_how_1964}. Thus, we do not make a specific recommendation regarding whether the physics GRE should be kept or removed as a result of our work because the answer to that question depends on the priorities of the department. However, if departments are using the physics GRE to identify applicants who might be missed by other metrics to achieve their admissions priorities, we suggest against this practice as it does not appear to be backed by evidence.

\acknowledgments{We would like to thank Julie Posselt, Casey Miller, and the Inclusive Graduate Network for providing us with the data for this project. We would also like to thank Rachel Henderson for her feedback on this project and guidance on the mediation and moderation analysis. This project was supported by the Michigan State University College of Natural Sciences and the Lappan-Phillips Foundation.}

\section{Appendix: Analysis of the Variables}\label{sec:appendix}

\begin{figure*}
 \includegraphics[width=0.8\textwidth]{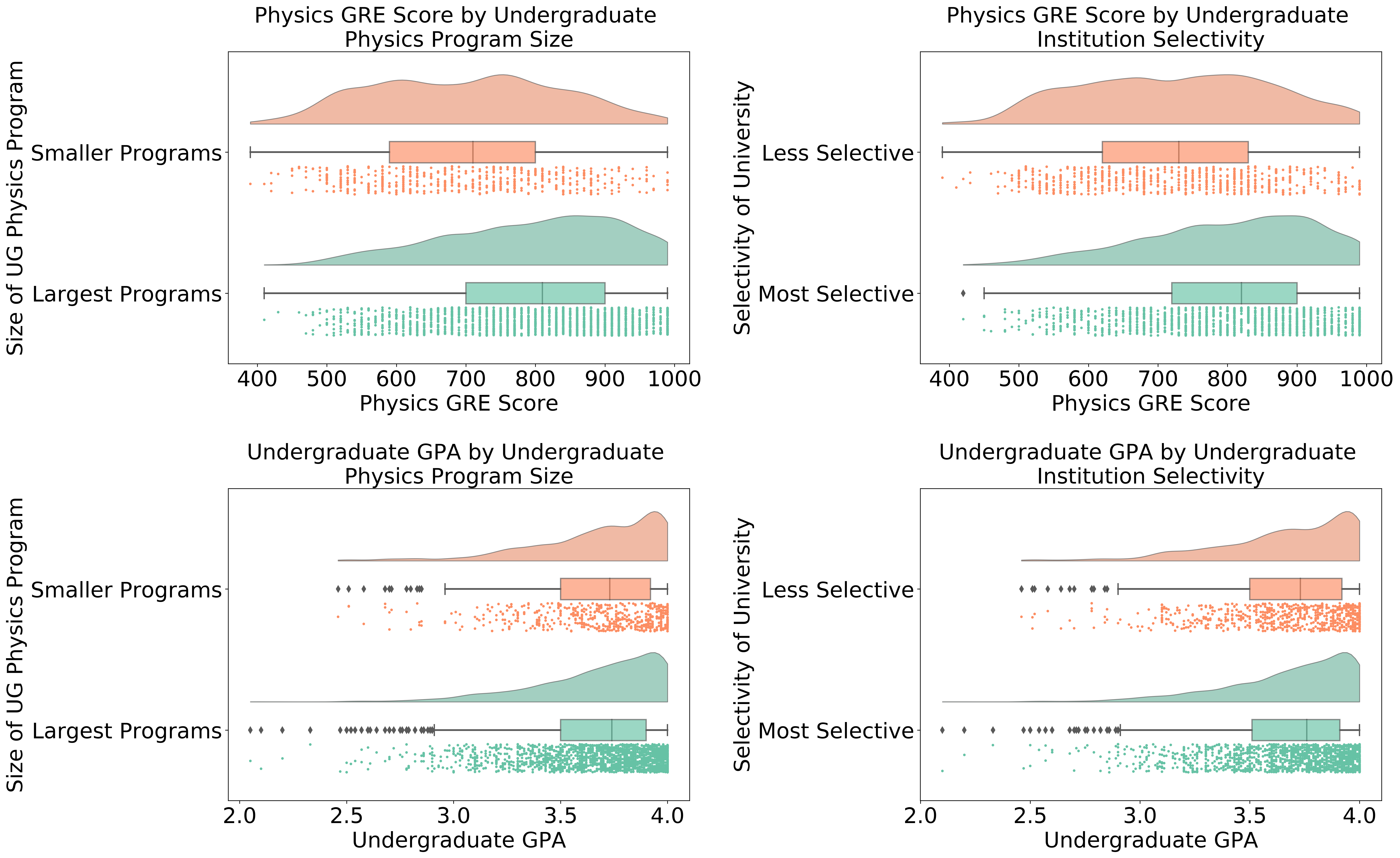}
 \caption{Distribution of physics GRE scores and undergraduate GPAs by the size of the undergraduate physics program and institutional selectivity for each applicant.\label{rain_inst}}
\end{figure*}

\begin{figure*}
 \includegraphics[width=0.8\textwidth]{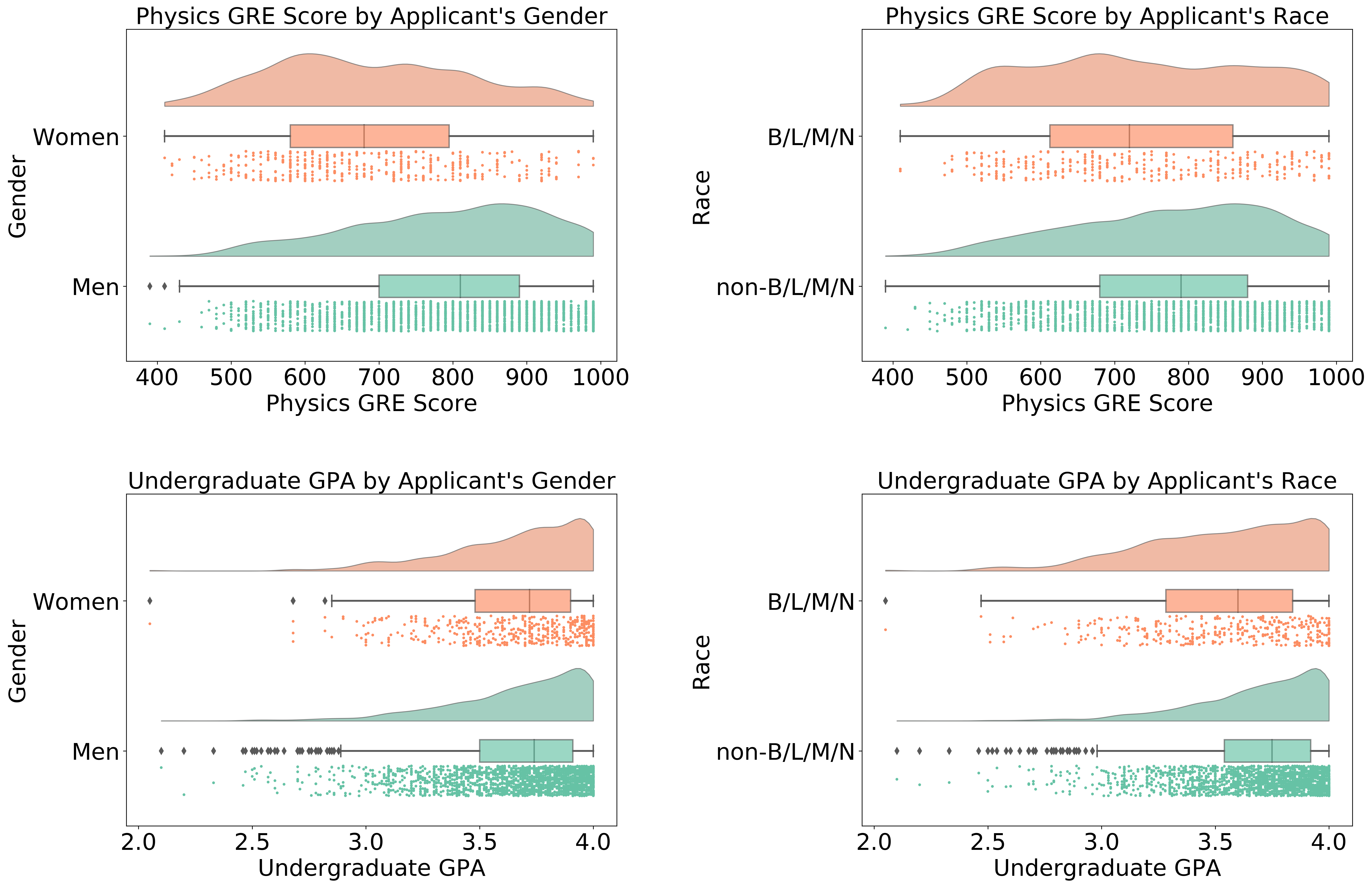}
 \caption{Distribution of physics GRE scores and undergraduate GPAs by gender and whether the applicant identified as a member of racial or ethnic group currently underrepresented in physics.\label{rain_demo}}
\end{figure*}

In this appendix, we describe the data used to answer research questions 2 and 3 to give the reader a better idea of what the distributions of physics GRE scores and GPA in the data set. Since the data are skewed left and exhibit ceiling effects (many applicants have 4.0 GPAs or 990 physics GRE scores), quartiles are used to describe the various features. To maximize the amount of information shown about the data, we use raincloud plots \cite{allen_raincloud_2019,whitaker_raincloudplotsraincloudplots_2019}, which show the distribution, the density plot, and traditional box plot. Kolmogorov-Smirnov tests suggest the distributions are not significantly different whether we include applicants who may have applied to multiple schools in our data set, so we include possible duplicates in our analysis.

Fig. \ref{rain_inst} shows the physics GRE scores and undergraduate GPAs of each applicant based on whether they attended a large undergraduate physics program (top 25\% nationally in yearly physics bachelor`s degrees) or attended a selective university (categorized as most competitive or highly competitive based on Barron`s Selectivity Index). We notice that the physics GRE score distributions are shifted to the right for applicants from large physics departments or selective institutions, signifying higher scores. Indeed, the median physics GRE scores of applicants from large programs or selective institutions is nearly 100 points higher than that of applicants from smaller or less selective institutions. However, in terms, of GPA, the median GPA is approximately the same, regardless of whether the applicant graduated from a larger or smaller physics department or attended a more or less selective institution.

Fig. \ref{rain_demo} shows the physics GRE and undergraduate GPAs by gender and race. As expected, men score higher on the physics GRE than women do and Asian and white applicants score higher than Black, Latinx, Multiracial, or Native applicants, though the gaps appear larger than those reported in \cite{miller_typical_2019}.

When comparing GPAs, we find that men and women have similar GPAs, as recently reported in \cite{whitcomb_not_2020} when comparing men and women's STEM GPAs. Likewise, our data also shows a racial GPA gap with non-B/L/M/N applicants having a median GPA higher than that of B/L/M/N applicants by 0.15.

When looking across both figures, we notice that the physics GRE score distributions from smaller and less selective programs resembles the physics GRE distributions of women and B/L/M/N applicants while the physics GRE score distributions of the largest and most selective programs resembles the physics GRE score distributions of men and non-B/L/M/N applicants. To see if gender and race are confounding variables in our analysis, we examined the fraction of women and B/L/M/N applicants in each group. If this were the case, the smaller and less selective programs should have a greater fraction of women and B/L/M/N applicants than the larger and more selective programs.

However, we did not find this to be the case. Applicants from more selective institutions were 16\% women while applicants from less selective institutions were 18\% women (15\% and 14\% respectively for B/L/M/N applicants). For institution size, applicants from larger institutions were 16\% women compared to 21\% women from smaller institutions (14\% and 17\% for B/L/M/N applicants respectively). Thus, it does not appear that differences in who attends (in terms of gender and race) larger or more selective institutions is responsible for the observed differences in scores.

\bibliography{Grad_school.bib} 

\begin{thebibliography}{48}%
\makeatletter
\providecommand \@ifxundefined [1]{%
 \@ifx{#1\undefined}
}%
\providecommand \@ifnum [1]{%
 \ifnum #1\expandafter \@firstoftwo
 \else \expandafter \@secondoftwo
 \fi
}%
\providecommand \@ifx [1]{%
 \ifx #1\expandafter \@firstoftwo
 \else \expandafter \@secondoftwo
 \fi
}%
\providecommand \natexlab [1]{#1}%
\providecommand \enquote  [1]{``#1''}%
\providecommand \bibnamefont  [1]{#1}%
\providecommand \bibfnamefont [1]{#1}%
\providecommand \citenamefont [1]{#1}%
\providecommand \href@noop [0]{\@secondoftwo}%
\providecommand \href [0]{\begingroup \@sanitize@url \@href}%
\providecommand \@href[1]{\@@startlink{#1}\@@href}%
\providecommand \@@href[1]{\endgroup#1\@@endlink}%
\providecommand \@sanitize@url [0]{\catcode `\\12\catcode `\$12\catcode
  `\&12\catcode `\#12\catcode `\^12\catcode `\_12\catcode `\%12\relax}%
\providecommand \@@startlink[1]{}%
\providecommand \@@endlink[0]{}%
\providecommand \url  [0]{\begingroup\@sanitize@url \@url }%
\providecommand \@url [1]{\endgroup\@href {#1}{\urlprefix }}%
\providecommand \urlprefix  [0]{URL }%
\providecommand \Eprint [0]{\href }%
\providecommand \doibase [0]{http://dx.doi.org/}%
\providecommand \selectlanguage [0]{\@gobble}%
\providecommand \bibinfo  [0]{\@secondoftwo}%
\providecommand \bibfield  [0]{\@secondoftwo}%
\providecommand \translation [1]{[#1]}%
\providecommand \BibitemOpen [0]{}%
\providecommand \bibitemStop [0]{}%
\providecommand \bibitemNoStop [0]{.\EOS\space}%
\providecommand \EOS [0]{\spacefactor3000\relax}%
\providecommand \BibitemShut  [1]{\csname bibitem#1\endcsname}%
\let\auto@bib@innerbib\@empty
\bibitem [{\citenamefont {Chari}\ and\ \citenamefont
  {Potvin}(2019)}]{chari_understanding_2019}%
  \BibitemOpen
  \bibfield  {author} {\bibinfo {author} {\bibfnamefont {D.}~\bibnamefont
  {Chari}}\ and\ \bibinfo {author} {\bibfnamefont {G.}~\bibnamefont {Potvin}},\
  }\href {\doibase 10.1103/PhysRevPhysEducRes.15.023101} {\bibfield  {journal}
  {\bibinfo  {journal} {Physical Review Physics Education Research}\ }\textbf
  {\bibinfo {volume} {15}},\ \bibinfo {pages} {023101} (\bibinfo {year}
  {2019})}\BibitemShut {NoStop}%
\bibitem [{\citenamefont {Potvin}\ \emph {et~al.}(2017)\citenamefont {Potvin},
  \citenamefont {Chari},\ and\ \citenamefont
  {Hodapp}}]{potvin_investigating_2017}%
  \BibitemOpen
  \bibfield  {author} {\bibinfo {author} {\bibfnamefont {G.}~\bibnamefont
  {Potvin}}, \bibinfo {author} {\bibfnamefont {D.}~\bibnamefont {Chari}}, \
  and\ \bibinfo {author} {\bibfnamefont {T.}~\bibnamefont {Hodapp}},\ }\href
  {\doibase 10.1103/PhysRevPhysEducRes.13.020142} {\bibfield  {journal}
  {\bibinfo  {journal} {Physical Review Physics Education Research}\ }\textbf
  {\bibinfo {volume} {13}},\ \bibinfo {pages} {020142} (\bibinfo {year}
  {2017})}\BibitemShut {NoStop}%
\bibitem [{\citenamefont {Posselt}(2016)}]{posselt_inside_2016}%
  \BibitemOpen
  \bibfield  {author} {\bibinfo {author} {\bibfnamefont {J.~R.}\ \bibnamefont
  {Posselt}},\ }\href@noop {} {\emph {\bibinfo {title} {Inside graduate
  admissions}}}\ (\bibinfo  {publisher} {Harvard University Press},\ \bibinfo
  {year} {2016})\BibitemShut {NoStop}%
\bibitem [{\citenamefont {Young}\ and\ \citenamefont
  {Caballero}(2020)}]{young_using_2020}%
  \BibitemOpen
  \bibfield  {author} {\bibinfo {author} {\bibfnamefont {N.~T.}\ \bibnamefont
  {Young}}\ and\ \bibinfo {author} {\bibfnamefont {M.~D.}\ \bibnamefont
  {Caballero}},\ }in\ \href {\doibase 10.1119/perc.2019.pr.Young} {\emph
  {\bibinfo {booktitle} {2019 {Physics} {Education} {Research} {Conference}
  {Proceedings}}}}\ (\bibinfo  {publisher} {American Association of Physics
  Teachers},\ \bibinfo {address} {Provo, UT},\ \bibinfo {year}
  {2020})\BibitemShut {NoStop}%
\bibitem [{\citenamefont {Miller}\ and\ \citenamefont
  {Stassun}(2014)}]{miller_test_2014}%
  \BibitemOpen
  \bibfield  {author} {\bibinfo {author} {\bibfnamefont {C.}~\bibnamefont
  {Miller}}\ and\ \bibinfo {author} {\bibfnamefont {K.}~\bibnamefont
  {Stassun}},\ }\href {\doibase 10.1038/nj7504-303a} {\bibfield  {journal}
  {\bibinfo  {journal} {Nature}\ }\textbf {\bibinfo {volume} {510}},\ \bibinfo
  {pages} {303} (\bibinfo {year} {2014})}\BibitemShut {NoStop}%
\bibitem [{\citenamefont {Miller}\ \emph {et~al.}(2019)\citenamefont {Miller},
  \citenamefont {Zwickl}, \citenamefont {Posselt}, \citenamefont
  {Silvestrini},\ and\ \citenamefont {Hodapp}}]{miller_typical_2019}%
  \BibitemOpen
  \bibfield  {author} {\bibinfo {author} {\bibfnamefont {C.~W.}\ \bibnamefont
  {Miller}}, \bibinfo {author} {\bibfnamefont {B.~M.}\ \bibnamefont {Zwickl}},
  \bibinfo {author} {\bibfnamefont {J.~R.}\ \bibnamefont {Posselt}}, \bibinfo
  {author} {\bibfnamefont {R.~T.}\ \bibnamefont {Silvestrini}}, \ and\ \bibinfo
  {author} {\bibfnamefont {T.}~\bibnamefont {Hodapp}},\ }\href {\doibase
  10.1126/sciadv.aat7550} {\bibfield  {journal} {\bibinfo  {journal} {Science
  Advances}\ }\textbf {\bibinfo {volume} {5}},\ \bibinfo {pages} {eaat7550}
  (\bibinfo {year} {2019})}\BibitemShut {NoStop}%
\bibitem [{\citenamefont {Cochran}\ \emph {et~al.}(2018)\citenamefont
  {Cochran}, \citenamefont {Hodapp},\ and\ \citenamefont
  {Brown}}]{cochran_identifying_2018}%
  \BibitemOpen
  \bibfield  {author} {\bibinfo {author} {\bibfnamefont {G.~L.}\ \bibnamefont
  {Cochran}}, \bibinfo {author} {\bibfnamefont {T.}~\bibnamefont {Hodapp}}, \
  and\ \bibinfo {author} {\bibfnamefont {E.~E.~A.}\ \bibnamefont {Brown}},\
  }in\ \href {\doibase http://dx.doi.org/10.1119/perc.2017.pr.018} {\emph
  {\bibinfo {booktitle} {Physics {Education} {Research} {Conference}
  {Proceedings}}}},\ \bibinfo {series and number} {{PER} {Conference}}\
  (\bibinfo {address} {Cincinnati, OH},\ \bibinfo {year} {2018})\ pp.\ \bibinfo
  {pages} {92--95}\BibitemShut {NoStop}%
\bibitem [{\citenamefont {Wilson}(2020)}]{wilson_predicting_2020}%
  \BibitemOpen
  \bibfield  {author} {\bibinfo {author} {\bibfnamefont {R.}~\bibnamefont
  {Wilson}},\ }\href {https://digitalcommons.kennesaw.edu/seceddoc_etd/22}
  {\bibfield  {journal} {\bibinfo  {journal} {Doctor of Education in Secondary
  Education Dissertations}\ } (\bibinfo {year} {2020})}\BibitemShut {NoStop}%
\bibitem [{\citenamefont {Owens}\ \emph {et~al.}(2020)\citenamefont {Owens},
  \citenamefont {Zwickl},\ and\ \citenamefont {Miller}}]{owens_not_2020}%
  \BibitemOpen
  \bibfield  {author} {\bibinfo {author} {\bibfnamefont {L.}~\bibnamefont
  {Owens}}, \bibinfo {author} {\bibfnamefont {B.}~\bibnamefont {Zwickl}}, \
  and\ \bibinfo {author} {\bibfnamefont {C.}~\bibnamefont {Miller}},\
  }\href@noop {} {\enquote {\bibinfo {title} {“{Not} {Required},”“{No}
  {Required} {Minimum},” and “{Optional}” {General} and {Physics} {GRE}
  {Requirements}: {The} {Impact} on {Prospective} {Graduate} {Students}},}\ }
  (\bibinfo {year} {2020}),\ \bibinfo {note} {presented at APS March
  Meeting}\BibitemShut {NoStop}%
\bibitem [{\citenamefont {Levesque}\ \emph {et~al.}(2015)\citenamefont
  {Levesque}, \citenamefont {Bezanson},\ and\ \citenamefont
  {Tremblay}}]{levesque_physics_2015}%
  \BibitemOpen
  \bibfield  {author} {\bibinfo {author} {\bibfnamefont {E.~M.}\ \bibnamefont
  {Levesque}}, \bibinfo {author} {\bibfnamefont {R.}~\bibnamefont {Bezanson}},
  \ and\ \bibinfo {author} {\bibfnamefont {G.~R.}\ \bibnamefont {Tremblay}},\
  }\href {http://arxiv.org/abs/1512.03709} {\bibfield  {journal} {\bibinfo
  {journal} {arXiv:1512.03709 [astro-ph, physics:physics]}\ } (\bibinfo {year}
  {2015})},\ \bibinfo {note} {arXiv: 1512.03709}\BibitemShut {NoStop}%
\bibitem [{noa(2019{\natexlab{a}})}]{noauthor_statement_2019}%
  \BibitemOpen
  \href
  {https://www.aapt.org/Resources/policy/upload/Statement_on_Use_of_GRE_for_Admission_to_Graduate_Physics_Programs.pdf}
  {\bibfield  {journal} {\bibinfo  {journal} {American Association of Physics
  Teacher}\ } (\bibinfo {year} {2019}{\natexlab{a}})}\BibitemShut {NoStop}%
\bibitem [{noa(2018)}]{noauthor_aas_2018}%
  \BibitemOpen
  \href {https://aas.org/about/governance/society-resolutions#GRE} {\enquote
  {\bibinfo {title} {{AAS} {Statement} on {Limiting} the {Use} of {GRE}
  {Scores} in {Graduate} {Admissions} in the {Astronomical} {Sciences}},}\ }
  (\bibinfo {year} {2018})\BibitemShut {NoStop}%
\bibitem [{\citenamefont {Lopez}(2019)}]{lopez_demographic_2019}%
  \BibitemOpen
  \bibfield  {author} {\bibinfo {author} {\bibfnamefont {L.~A.}\ \bibnamefont
  {Lopez}},\ }\href@noop {} {\enquote {\bibinfo {title} {Demographic {Effects}
  of {Removing} the {Physics} {GRE} {Requirement} in {Graduate}
  {Admissions}},}\ } (\bibinfo {year} {2019})\BibitemShut {NoStop}%
\bibitem [{\citenamefont {Scherr}\ \emph {et~al.}(2017)\citenamefont {Scherr},
  \citenamefont {Plisch}, \citenamefont {Gray}, \citenamefont {Potvin},\ and\
  \citenamefont {Hodapp}}]{scherr_fixed_2017}%
  \BibitemOpen
  \bibfield  {author} {\bibinfo {author} {\bibfnamefont {R.~E.}\ \bibnamefont
  {Scherr}}, \bibinfo {author} {\bibfnamefont {M.}~\bibnamefont {Plisch}},
  \bibinfo {author} {\bibfnamefont {K.~E.}\ \bibnamefont {Gray}}, \bibinfo
  {author} {\bibfnamefont {G.}~\bibnamefont {Potvin}}, \ and\ \bibinfo {author}
  {\bibfnamefont {T.}~\bibnamefont {Hodapp}},\ }\href {\doibase
  10.1103/PhysRevPhysEducRes.13.020133} {\bibfield  {journal} {\bibinfo
  {journal} {Physical Review Physics Education Research}\ }\textbf {\bibinfo
  {volume} {13}},\ \bibinfo {pages} {020133} (\bibinfo {year} {2017})},\
  \bibinfo {note} {publisher: American Physical Society}\BibitemShut {NoStop}%
\bibitem [{\citenamefont {Leslie}\ \emph {et~al.}(2015)\citenamefont {Leslie},
  \citenamefont {Cimpian}, \citenamefont {Meyer},\ and\ \citenamefont
  {Freeland}}]{leslie_expectations_2015}%
  \BibitemOpen
  \bibfield  {author} {\bibinfo {author} {\bibfnamefont {S.-J.}\ \bibnamefont
  {Leslie}}, \bibinfo {author} {\bibfnamefont {A.}~\bibnamefont {Cimpian}},
  \bibinfo {author} {\bibfnamefont {M.}~\bibnamefont {Meyer}}, \ and\ \bibinfo
  {author} {\bibfnamefont {E.}~\bibnamefont {Freeland}},\ }\href {\doibase
  10.1126/science.1261375} {\bibfield  {journal} {\bibinfo  {journal}
  {Science}\ }\textbf {\bibinfo {volume} {347}},\ \bibinfo {pages} {262}
  (\bibinfo {year} {2015})},\ \bibinfo {note} {publisher: American Association
  for the Advancement of Science Section: Report}\BibitemShut {NoStop}%
\bibitem [{\citenamefont {Langin}(2019)}]{langin_wave_2019}%
  \BibitemOpen
  \bibfield  {author} {\bibinfo {author} {\bibfnamefont {K.}~\bibnamefont
  {Langin}},\ }\href {\doibase 10.1126/science.caredit.aay2093} {\bibfield
  {journal} {\bibinfo  {journal} {Science}\ } (\bibinfo {year} {2019}),\
  10.1126/science.caredit.aay2093}\BibitemShut {NoStop}%
\bibitem [{noa()}]{noauthor_about_nodate}%
  \BibitemOpen
  \href {https://www.ets.org/gre/subject/about} {\enquote {\bibinfo {title}
  {About the {GRE} {Subject} {Tests} ({For} {Test} {Takers})},}\ }\BibitemShut
  {NoStop}%
\bibitem [{noa(2019{\natexlab{b}})}]{noauthor_gre_2019}%
  \BibitemOpen
  \href {https://www.ets.org/s/gre/pdf/gre_guide.pdf} {\enquote {\bibinfo
  {title} {{GRE} {Guide} to the {Use} of {Scores}},}\ } (\bibinfo {year}
  {2019}{\natexlab{b}})\BibitemShut {NoStop}%
\bibitem [{\citenamefont {Morrison}\ \emph {et~al.}(2020)\citenamefont
  {Morrison}, \citenamefont {Dixon}, \citenamefont {Miller},\ and\
  \citenamefont {in~Astronomy IV Graduate School Admissions White
  Paper~Group}}]{morrison_women_2020}%
  \BibitemOpen
  \bibfield  {author} {\bibinfo {author} {\bibfnamefont {N.~D.}\ \bibnamefont
  {Morrison}}, \bibinfo {author} {\bibfnamefont {W.~V.}\ \bibnamefont {Dixon}},
  \bibinfo {author} {\bibfnamefont {C.~W.}\ \bibnamefont {Miller}}, \ and\
  \bibinfo {author} {\bibfnamefont {T.~W.}\ \bibnamefont {in~Astronomy IV
  Graduate School Admissions White Paper~Group}},\ }\href
  {https://baas.aas.org/pub/2019i0204/release/1} {\bibfield  {journal}
  {\bibinfo  {journal} {Bulletin of the AAS}\ }\textbf {\bibinfo {volume} {51}}
  (\bibinfo {year} {2020})}\BibitemShut {NoStop}%
\bibitem [{\citenamefont {Levesque}\ \emph {et~al.}(2017)\citenamefont
  {Levesque}, \citenamefont {Bezanson},\ and\ \citenamefont
  {Tremblay}}]{levesque_why_2017}%
  \BibitemOpen
  \bibfield  {author} {\bibinfo {author} {\bibfnamefont {E.~M.}\ \bibnamefont
  {Levesque}}, \bibinfo {author} {\bibfnamefont {R.}~\bibnamefont {Bezanson}},
  \ and\ \bibinfo {author} {\bibfnamefont {G.~R.}\ \bibnamefont {Tremblay}},\
  }\href {\doibase 10.1063/PT.5.9090} {\bibfield  {journal} {\bibinfo
  {journal} {Physics Today}\ } (\bibinfo {year} {2017}),\ 10.1063/PT.5.9090},\
  \bibinfo {note} {publisher: American Institute of Physics}\BibitemShut
  {NoStop}%
\bibitem [{\citenamefont {Posselt}(2018)}]{posselt_trust_2018}%
  \BibitemOpen
  \bibfield  {author} {\bibinfo {author} {\bibfnamefont {J.~R.}\ \bibnamefont
  {Posselt}},\ }\href {\doibase 10.1353/rhe.2018.0023} {\bibfield  {journal}
  {\bibinfo  {journal} {The Review of Higher Education}\ }\textbf {\bibinfo
  {volume} {41}},\ \bibinfo {pages} {497} (\bibinfo {year} {2018})}\BibitemShut
  {NoStop}%
\bibitem [{\citenamefont {Small}(2017)}]{small_range_2017}%
  \BibitemOpen
  \bibfield  {author} {\bibinfo {author} {\bibfnamefont {A.~R.}\ \bibnamefont
  {Small}},\ }\href {http://arxiv.org/abs/1709.02895} {\bibfield  {journal}
  {\bibinfo  {journal} {arXiv:1709.02895 [physics]}\ } (\bibinfo {year}
  {2017})},\ \bibinfo {note} {arXiv: 1709.02895}\BibitemShut {NoStop}%
\bibitem [{\citenamefont {Baron}\ and\ \citenamefont
  {Kenny}(1986)}]{baron_moderator-mediator_1986}%
  \BibitemOpen
  \bibfield  {author} {\bibinfo {author} {\bibfnamefont {R.~M.}\ \bibnamefont
  {Baron}}\ and\ \bibinfo {author} {\bibfnamefont {D.~A.}\ \bibnamefont
  {Kenny}},\ }\href@noop {} {\bibfield  {journal} {\bibinfo  {journal} {Journal
  of Personality and Social Psychology}\ }\textbf {\bibinfo {volume} {51}},\
  \bibinfo {pages} {1173} (\bibinfo {year} {1986})}\BibitemShut {NoStop}%
\bibitem [{\citenamefont {Rijnhart}\ \emph {et~al.}(2019)\citenamefont
  {Rijnhart}, \citenamefont {Twisk}, \citenamefont {Eekhout},\ and\
  \citenamefont {Heymans}}]{rijnhart_comparison_2019}%
  \BibitemOpen
  \bibfield  {author} {\bibinfo {author} {\bibfnamefont {J.~J.~M.}\
  \bibnamefont {Rijnhart}}, \bibinfo {author} {\bibfnamefont {J.~W.~R.}\
  \bibnamefont {Twisk}}, \bibinfo {author} {\bibfnamefont {I.}~\bibnamefont
  {Eekhout}}, \ and\ \bibinfo {author} {\bibfnamefont {M.~W.}\ \bibnamefont
  {Heymans}},\ }\href {\doibase 10.1186/s12874-018-0654-z} {\bibfield
  {journal} {\bibinfo  {journal} {BMC Medical Research Methodology}\ }\textbf
  {\bibinfo {volume} {19}},\ \bibinfo {pages} {19} (\bibinfo {year}
  {2019})}\BibitemShut {NoStop}%
\bibitem [{\citenamefont {Hayes}\ and\ \citenamefont
  {Scharkow}(2013)}]{hayes_relative_2013}%
  \BibitemOpen
  \bibfield  {author} {\bibinfo {author} {\bibfnamefont {A.~F.}\ \bibnamefont
  {Hayes}}\ and\ \bibinfo {author} {\bibfnamefont {M.}~\bibnamefont
  {Scharkow}},\ }\href {\doibase 10.1177/0956797613480187} {\bibfield
  {journal} {\bibinfo  {journal} {Psychological Science}\ }\textbf {\bibinfo
  {volume} {24}},\ \bibinfo {pages} {1918} (\bibinfo {year} {2013})},\ \bibinfo
  {note} {publisher: SAGE Publications Inc}\BibitemShut {NoStop}%
\bibitem [{\citenamefont {Amos}\ and\ \citenamefont
  {Heckler}(2018)}]{amos_mediating_2018}%
  \BibitemOpen
  \bibfield  {author} {\bibinfo {author} {\bibfnamefont {N.}~\bibnamefont
  {Amos}}\ and\ \bibinfo {author} {\bibfnamefont {A.~F.}\ \bibnamefont
  {Heckler}},\ }\href {\doibase 10.1103/PhysRevPhysEducRes.14.010105}
  {\bibfield  {journal} {\bibinfo  {journal} {Physical Review Physics Education
  Research}\ }\textbf {\bibinfo {volume} {14}},\ \bibinfo {pages} {010105}
  (\bibinfo {year} {2018})},\ \bibinfo {note} {publisher: American Physical
  Society}\BibitemShut {NoStop}%
\bibitem [{\citenamefont {Ditlevsen}\ \emph {et~al.}(2005)\citenamefont
  {Ditlevsen}, \citenamefont {Christensen}, \citenamefont {Lynch},
  \citenamefont {Damsgaard},\ and\ \citenamefont
  {Keiding}}]{ditlevsen_mediation_2005}%
  \BibitemOpen
  \bibfield  {author} {\bibinfo {author} {\bibfnamefont {S.}~\bibnamefont
  {Ditlevsen}}, \bibinfo {author} {\bibfnamefont {U.}~\bibnamefont
  {Christensen}}, \bibinfo {author} {\bibfnamefont {J.}~\bibnamefont {Lynch}},
  \bibinfo {author} {\bibfnamefont {M.~T.}\ \bibnamefont {Damsgaard}}, \ and\
  \bibinfo {author} {\bibfnamefont {N.}~\bibnamefont {Keiding}},\ }\href
  {\doibase 10.1097/01.ede.0000147107.76079.07} {\bibfield  {journal} {\bibinfo
   {journal} {Epidemiology}\ }\textbf {\bibinfo {volume} {16}},\ \bibinfo
  {pages} {114} (\bibinfo {year} {2005})}\BibitemShut {NoStop}%
\bibitem [{\citenamefont {Freedman}(2001)}]{freedman_confidence_2001}%
  \BibitemOpen
  \bibfield  {author} {\bibinfo {author} {\bibfnamefont {L.~S.}\ \bibnamefont
  {Freedman}},\ }\href {\doibase 10.1016/S0378-3758(00)00330-X} {\bibfield
  {journal} {\bibinfo  {journal} {Journal of Statistical Planning and
  Inference}\ }\textbf {\bibinfo {volume} {96}},\ \bibinfo {pages} {143}
  (\bibinfo {year} {2001})}\BibitemShut {NoStop}%
\bibitem [{\citenamefont {Preacher}\ \emph {et~al.}(2007)\citenamefont
  {Preacher}, \citenamefont {Rucker},\ and\ \citenamefont
  {Hayes}}]{preacher_addressing_2007}%
  \BibitemOpen
  \bibfield  {author} {\bibinfo {author} {\bibfnamefont {K.~J.}\ \bibnamefont
  {Preacher}}, \bibinfo {author} {\bibfnamefont {D.~D.}\ \bibnamefont
  {Rucker}}, \ and\ \bibinfo {author} {\bibfnamefont {A.~F.}\ \bibnamefont
  {Hayes}},\ }\href {\doibase 10.1080/00273170701341316} {\bibfield  {journal}
  {\bibinfo  {journal} {Multivariate Behavioral Research}\ }\textbf {\bibinfo
  {volume} {42}},\ \bibinfo {pages} {185} (\bibinfo {year} {2007})}\BibitemShut
  {NoStop}%
\bibitem [{\citenamefont {Weissman}(2020)}]{weissman_gre_2020}%
  \BibitemOpen
  \bibfield  {author} {\bibinfo {author} {\bibfnamefont {M.~B.}\ \bibnamefont
  {Weissman}},\ }\href {\doibase 10.1126/sciadv.aax3787} {\bibfield  {journal}
  {\bibinfo  {journal} {Science Advances}\ }\textbf {\bibinfo {volume} {6}},\
  \bibinfo {pages} {eaax3787} (\bibinfo {year} {2020})},\ \bibinfo {note}
  {publisher: American Association for the Advancement of Science Section:
  Technical Comments}\BibitemShut {NoStop}%
\bibitem [{\citenamefont {Nicholson}\ and\ \citenamefont
  {Mulvey}(2018)}]{nicholson_roster_2018}%
  \BibitemOpen
  \bibfield  {author} {\bibinfo {author} {\bibfnamefont {S.}~\bibnamefont
  {Nicholson}}\ and\ \bibinfo {author} {\bibfnamefont {P.~J.}\ \bibnamefont
  {Mulvey}},\ }\href
  {https://www.aip.org/statistics/reports/roster-physics-2017} {\emph {\bibinfo
  {title} {Roster of {Physics} {Departments} with {Enrollment} and {Degree}
  {Data}, 2017}}},\ \bibinfo {type} {Tech. Rep.}\ (\bibinfo  {institution}
  {American Institute of Physics},\ \bibinfo {year} {2018})\BibitemShut
  {NoStop}%
\bibitem [{\citenamefont {Nicholson}\ and\ \citenamefont
  {Mulvey}(2019)}]{nicholson_roster_2019}%
  \BibitemOpen
  \bibfield  {author} {\bibinfo {author} {\bibfnamefont {S.}~\bibnamefont
  {Nicholson}}\ and\ \bibinfo {author} {\bibfnamefont {P.~J.}\ \bibnamefont
  {Mulvey}},\ }\href
  {https://www.aip.org/statistics/reports/roster-physics-2018} {\emph {\bibinfo
  {title} {Roster of {Physics} {Departments} with {Enrollment} and {Degree}
  {Data}, 2018}}},\ \bibinfo {type} {Tech. Rep.}\ (\bibinfo  {institution}
  {American Institute of Physics},\ \bibinfo {year} {2019})\BibitemShut
  {NoStop}%
\bibitem [{\citenamefont {{National Center for Education
  Statistics}}(2017)}]{national_center_for_education_statistics_nces-barrons_2017}%
  \BibitemOpen
  \bibfield  {author} {\bibinfo {author} {\bibnamefont {{National Center for
  Education Statistics}}},\ }\href
  {https://nces.ed.gov/pubsearch/pubsinfo.asp?pubid=2016332} {\enquote
  {\bibinfo {title} {{NCES}-{Barron}'s {Admissions} {Competitiveness} {Index}
  {Data} {Files}: 1972, 1982, 1992, 2004, , 2008, 2014},}\ } (\bibinfo {year}
  {2017}),\ \bibinfo {note} {library Catalog: nces.ed.gov Publisher: National
  Center for Education Statistics}\BibitemShut {NoStop}%
\bibitem [{\citenamefont {Chetty}\ \emph {et~al.}(2017)\citenamefont {Chetty},
  \citenamefont {Friedman}, \citenamefont {Saez}, \citenamefont {Turner},\ and\
  \citenamefont {Yagan}}]{chetty_mobility_2017}%
  \BibitemOpen
  \bibfield  {author} {\bibinfo {author} {\bibfnamefont {R.}~\bibnamefont
  {Chetty}}, \bibinfo {author} {\bibfnamefont {J.}~\bibnamefont {Friedman}},
  \bibinfo {author} {\bibfnamefont {E.}~\bibnamefont {Saez}}, \bibinfo {author}
  {\bibfnamefont {N.}~\bibnamefont {Turner}}, \ and\ \bibinfo {author}
  {\bibfnamefont {D.}~\bibnamefont {Yagan}},\ }\href {\doibase 10.3386/w23618}
  {\emph {\bibinfo {title} {Mobility {Report} {Cards}: {The} {Role} of
  {Colleges} in {Intergenerational} {Mobility}}}},\ \bibinfo {type} {Tech.
  Rep.}\ \bibinfo {number} {w23618}\ (\bibinfo  {institution} {National Bureau
  of Economic Research},\ \bibinfo {address} {Cambridge, MA},\ \bibinfo {year}
  {2017})\BibitemShut {NoStop}%
\bibitem [{\citenamefont {Blue}\ \emph {et~al.}(2018)\citenamefont {Blue},
  \citenamefont {Traxler},\ and\ \citenamefont {Cid}}]{blue_gender_2018}%
  \BibitemOpen
  \bibfield  {author} {\bibinfo {author} {\bibfnamefont {J.}~\bibnamefont
  {Blue}}, \bibinfo {author} {\bibfnamefont {A.~L.}\ \bibnamefont {Traxler}}, \
  and\ \bibinfo {author} {\bibfnamefont {X.~C.}\ \bibnamefont {Cid}},\ }\href
  {\doibase 10.1063/PT.3.3870} {\bibfield  {journal} {\bibinfo  {journal}
  {Physics Today}\ }\textbf {\bibinfo {volume} {71}},\ \bibinfo {pages} {40}
  (\bibinfo {year} {2018})}\BibitemShut {NoStop}%
\bibitem [{\citenamefont {Williams}(2020)}]{williams_underrepresented_2020}%
  \BibitemOpen
  \bibfield  {author} {\bibinfo {author} {\bibfnamefont {T.~L.}\ \bibnamefont
  {Williams}},\ }\href
  {https://cacm.acm.org/blogs/blog-cacm/245710-underrepresented-minority-considered-harmful-racist-language/fulltext}
  {\enquote {\bibinfo {title} {'{Underrepresented} {Minority}' {Considered}
  {Harmful}, {Racist} {Language}},}\ } (\bibinfo {year} {2020}),\ \bibinfo
  {note} {library Catalog: cacm.acm.org}\BibitemShut {NoStop}%
\bibitem [{\citenamefont {Teranishi}(2007)}]{teranishi_race_2007}%
  \BibitemOpen
  \bibfield  {author} {\bibinfo {author} {\bibfnamefont {R.~T.}\ \bibnamefont
  {Teranishi}},\ }\href@noop {} {\bibfield  {journal} {\bibinfo  {journal} {New
  Directions for Institutional Research}\ }\textbf {\bibinfo {volume} {2007}},\
  \bibinfo {pages} {37} (\bibinfo {year} {2007})},\ \bibinfo {note} {publisher:
  Wiley Online Library}\BibitemShut {NoStop}%
\bibitem [{\citenamefont {Posselt}\ \emph {et~al.}(2019)\citenamefont
  {Posselt}, \citenamefont {Hernandez}, \citenamefont {Cochran},\ and\
  \citenamefont {Miller}}]{posselt_metrics_2019}%
  \BibitemOpen
  \bibfield  {author} {\bibinfo {author} {\bibfnamefont {J.}~\bibnamefont
  {Posselt}}, \bibinfo {author} {\bibfnamefont {T.}~\bibnamefont {Hernandez}},
  \bibinfo {author} {\bibfnamefont {G.}~\bibnamefont {Cochran}}, \ and\
  \bibinfo {author} {\bibfnamefont {C.}~\bibnamefont {Miller}},\ }\href
  {\doibase 10.1615/JWomenMinorScienEng.2019027863} {\bibfield  {journal}
  {\bibinfo  {journal} {Journal of Women and Minorities in Science and
  Engineering}\ }\textbf {\bibinfo {volume} {25}} (\bibinfo {year} {2019}),\
  10.1615/JWomenMinorScienEng.2019027863}\BibitemShut {NoStop}%
\bibitem [{\citenamefont
  {Massey~Jr.}(1951)}]{massey_jr_kolmogorov-smirnov_1951}%
  \BibitemOpen
  \bibfield  {author} {\bibinfo {author} {\bibfnamefont {F.~J.}\ \bibnamefont
  {Massey~Jr.}},\ }\href {\doibase 10.1080/01621459.1951.10500769} {\bibfield
  {journal} {\bibinfo  {journal} {Journal of the American Statistical
  Association}\ }\textbf {\bibinfo {volume} {46}},\ \bibinfo {pages} {68}
  (\bibinfo {year} {1951})}\BibitemShut {NoStop}%
\bibitem [{\citenamefont {Miller}\ and\ \citenamefont
  {Posselt}(2018)}]{miller_broadening_2018}%
  \BibitemOpen
  \bibfield  {author} {\bibinfo {author} {\bibfnamefont {C.}~\bibnamefont
  {Miller}}\ and\ \bibinfo {author} {\bibfnamefont {J.}~\bibnamefont
  {Posselt}},\ }\href {https://meetings.aps.org/Meeting/BPNMC18/Session/W3A.1}
  {\enquote {\bibinfo {title} {Broadening {Participation} in {Graduate}
  {Education} through {Holistic} {Review}},}\ } (\bibinfo {year} {2018}),\
  \bibinfo {note} {presented at the Bridge Program and National Mentoring
  Community Conference}\BibitemShut {NoStop}%
\bibitem [{\citenamefont {Kent}\ and\ \citenamefont
  {McCarthy}(2016)}]{kent_holistic_2016}%
  \BibitemOpen
  \bibfield  {author} {\bibinfo {author} {\bibfnamefont {J.~D.}\ \bibnamefont
  {Kent}}\ and\ \bibinfo {author} {\bibfnamefont {M.~T.}\ \bibnamefont
  {McCarthy}},\ }\href
  {https://cgsnet.org/ckfinder/userfiles/files/CGS_HolisticReview_final_web.pdf}
  {\bibfield  {journal} {\bibinfo  {journal} {Washington DC: Council of
  Graduate Students}\ } (\bibinfo {year} {2016})}\BibitemShut {NoStop}%
\bibitem [{\citenamefont {Rudolph}\ \emph {et~al.}(2020)\citenamefont
  {Rudolph}, \citenamefont {Basri}, \citenamefont {Agüeros}, \citenamefont
  {Bertschinger}, \citenamefont {Coble}, \citenamefont {Donahue}, \citenamefont
  {Monkiewicz}, \citenamefont {Speck}, \citenamefont {Stassun}, \citenamefont
  {Ivie}, \citenamefont {Pfund},\ and\ \citenamefont
  {Posselt}}]{rudolph_final_2020}%
  \BibitemOpen
  \bibfield  {author} {\bibinfo {author} {\bibfnamefont {A.}~\bibnamefont
  {Rudolph}}, \bibinfo {author} {\bibfnamefont {G.}~\bibnamefont {Basri}},
  \bibinfo {author} {\bibfnamefont {M.}~\bibnamefont {Agüeros}}, \bibinfo
  {author} {\bibfnamefont {E.}~\bibnamefont {Bertschinger}}, \bibinfo {author}
  {\bibfnamefont {K.}~\bibnamefont {Coble}}, \bibinfo {author} {\bibfnamefont
  {M.}~\bibnamefont {Donahue}}, \bibinfo {author} {\bibfnamefont
  {J.}~\bibnamefont {Monkiewicz}}, \bibinfo {author} {\bibfnamefont
  {A.}~\bibnamefont {Speck}}, \bibinfo {author} {\bibfnamefont
  {K.}~\bibnamefont {Stassun}}, \bibinfo {author} {\bibfnamefont
  {R.}~\bibnamefont {Ivie}}, \bibinfo {author} {\bibfnamefont {C.}~\bibnamefont
  {Pfund}}, \ and\ \bibinfo {author} {\bibfnamefont {J.}~\bibnamefont
  {Posselt}},\ }\href {https://baas.aas.org/pub/2019i0101} {\bibfield
  {journal} {\bibinfo  {journal} {Bulletin of the AAS}\ }\textbf {\bibinfo
  {volume} {51}} (\bibinfo {year} {2020})}\BibitemShut {NoStop}%
\bibitem [{\citenamefont {Dunlap}(2018)}]{dunlap_journey_2018}%
  \BibitemOpen
  \bibfield  {author} {\bibinfo {author} {\bibfnamefont {K.}~\bibnamefont
  {Dunlap}},\ }\href
  {https://cpb-us-w2.wpmucdn.com/u.osu.edu/dist/d/39848/files/2018/07/Kris-AGPA-preso-for-publication-abbrev-w45oyd.pdf}
  {\enquote {\bibinfo {title} {Journey to a more holistic admissions review
  process by implementing an evaluation rubric},}\ } (\bibinfo {year}
  {2018})\BibitemShut {NoStop}%
\bibitem [{\citenamefont {Wilson}\ \emph {et~al.}(2019)\citenamefont {Wilson},
  \citenamefont {Odem}, \citenamefont {Walters}, \citenamefont {DePass},\ and\
  \citenamefont {Bean}}]{wilson_model_2019}%
  \BibitemOpen
  \bibfield  {author} {\bibinfo {author} {\bibfnamefont {M.~A.}\ \bibnamefont
  {Wilson}}, \bibinfo {author} {\bibfnamefont {M.~A.}\ \bibnamefont {Odem}},
  \bibinfo {author} {\bibfnamefont {T.}~\bibnamefont {Walters}}, \bibinfo
  {author} {\bibfnamefont {A.~L.}\ \bibnamefont {DePass}}, \ and\ \bibinfo
  {author} {\bibfnamefont {A.~J.}\ \bibnamefont {Bean}},\ }\href {\doibase
  10.1187/cbe.18-06-0103} {\bibfield  {journal} {\bibinfo  {journal} {CBE Life
  Sciences Education}\ }\textbf {\bibinfo {volume} {18}} (\bibinfo {year}
  {2019}),\ 10.1187/cbe.18-06-0103}\BibitemShut {NoStop}%
\bibitem [{\citenamefont {Searle}(1964)}]{searle_how_1964}%
  \BibitemOpen
  \bibfield  {author} {\bibinfo {author} {\bibfnamefont {J.~R.}\ \bibnamefont
  {Searle}},\ }\href {\doibase 10.2307/2183201} {\bibfield  {journal} {\bibinfo
   {journal} {The Philosophical Review}\ }\textbf {\bibinfo {volume} {73}},\
  \bibinfo {pages} {43} (\bibinfo {year} {1964})},\ \bibinfo {note} {publisher:
  [Duke University Press, Philosophical Review]}\BibitemShut {NoStop}%
\bibitem [{\citenamefont {Allen}\ \emph {et~al.}(2019)\citenamefont {Allen},
  \citenamefont {Poggiali}, \citenamefont {Whitaker}, \citenamefont
  {Marshall},\ and\ \citenamefont {Kievit}}]{allen_raincloud_2019}%
  \BibitemOpen
  \bibfield  {author} {\bibinfo {author} {\bibfnamefont {M.}~\bibnamefont
  {Allen}}, \bibinfo {author} {\bibfnamefont {D.}~\bibnamefont {Poggiali}},
  \bibinfo {author} {\bibfnamefont {K.}~\bibnamefont {Whitaker}}, \bibinfo
  {author} {\bibfnamefont {T.~R.}\ \bibnamefont {Marshall}}, \ and\ \bibinfo
  {author} {\bibfnamefont {R.~A.}\ \bibnamefont {Kievit}},\ }\href {\doibase
  10.12688/wellcomeopenres.15191.1} {\bibfield  {journal} {\bibinfo  {journal}
  {Wellcome Open Research}\ }\textbf {\bibinfo {volume} {4}},\ \bibinfo {pages}
  {63} (\bibinfo {year} {2019})}\BibitemShut {NoStop}%
\bibitem [{\citenamefont {Whitaker}\ \emph {et~al.}(2019)\citenamefont
  {Whitaker}, \citenamefont {Marshall}, \citenamefont {Mourik}, \citenamefont
  {Martinez}, \citenamefont {Poggiali}, \citenamefont {Ye},\ and\ \citenamefont
  {Klug}}]{whitaker_raincloudplotsraincloudplots_2019}%
  \BibitemOpen
  \bibfield  {author} {\bibinfo {author} {\bibfnamefont {K.}~\bibnamefont
  {Whitaker}}, \bibinfo {author} {\bibfnamefont {T.~R.}\ \bibnamefont
  {Marshall}}, \bibinfo {author} {\bibfnamefont {T.~V.}\ \bibnamefont
  {Mourik}}, \bibinfo {author} {\bibfnamefont {P.~A.}\ \bibnamefont
  {Martinez}}, \bibinfo {author} {\bibfnamefont {D.}~\bibnamefont {Poggiali}},
  \bibinfo {author} {\bibfnamefont {H.}~\bibnamefont {Ye}}, \ and\ \bibinfo
  {author} {\bibfnamefont {M.}~\bibnamefont {Klug}},\ }\href {\doibase
  10.5281/ZENODO.3368186} {\enquote {\bibinfo {title}
  {{RainCloudPlots}/{RainCloudPlots}: {WellcomeOpenResearch}},}\ } (\bibinfo
  {year} {2019})\BibitemShut {NoStop}%
\bibitem [{\citenamefont {Whitcomb}\ and\ \citenamefont
  {Singh}(2020)}]{whitcomb_not_2020}%
  \BibitemOpen
  \bibfield  {author} {\bibinfo {author} {\bibfnamefont {K.~M.}\ \bibnamefont
  {Whitcomb}}\ and\ \bibinfo {author} {\bibfnamefont {C.}~\bibnamefont
  {Singh}},\ }\href {http://arxiv.org/abs/2003.04376} {\bibfield  {journal}
  {\bibinfo  {journal} {arXiv:2003.04376 [physics]}\ } (\bibinfo {year}
  {2020})},\ \bibinfo {note} {arXiv: 2003.04376}\BibitemShut {NoStop}%
\end{thebibliography}%

\end{document}